\documentclass[preprint,amsmath]{revtex4-1}

\usepackage{epsfig}
\usepackage{amsmath}
\usepackage{amssymb}
\usepackage{color}
\usepackage[normalem]{ulem}
\usepackage{caption}
\usepackage{subcaption}

\begin{document}

\title{The Noisy Oscillator : Random Mass and Random Damping }
\author{Stanislav Burov}
\email{stasbur@gmail.com}
\author{Moshe Gitterman}
\email{moshe.gitterman@biu.ac.il }
\affiliation{Physics Department, Bar-Ilan University, Ramat Gan 52900,
Israel}
\pacs{PACS}

\begin{abstract}
The problem of a linear damped noisy oscillator is treated in the presence
of two multiplicative sources of noise which imply a random mass and random
damping. The additive noise and the noise in the damping are responsible for
an influx of energy to the oscillator and its dissipation to the surrounding
environment. A random mass implies that the surrounding molecules not only
collide with the oscillator but may also adhere to it, thereby changing its
mass. We present general formulas for the first two moments and address the
question of mean and energetic stabilities. The phenomenon of stochastic
resonance, i.e. the expansion due to the noise of a system response to an
external periodic signal, is considered for separate and joint action of two
sources of noise and their characteristics.
\end{abstract}

\maketitle

\section{Introduction}

One of the most general and most widely used models in physics is the damped
linear harmonic oscillator, which is described by the following equation 
\begin{equation}
m\frac{d^{2}x}{dt^{2}}+\gamma \frac{dx}{dt}+\omega ^{2}x=0  \label{int01}
\end{equation}%
%
This model has been applied in many fields, ranging from quarcks to
cosmology. The ancient Greeks already had a general idea of oscillations and
used them in musical instruments. Many applications have been found in the
last 400 years~\cite{GitBook}. The solution of Eq. (\ref{int01}) depends on
the parameters $\gamma /m$ and $\omega ^{2}/m.$ For a solution of the type $%
x=\exp \left( \alpha t\right) $, one obtains $\alpha =-\frac{\gamma }{2m}\pm 
\sqrt{\frac{\gamma ^{2}}{4m^{2}}-}\frac{\omega ^{2}}{m}.$ For $\left( \gamma
/m\right) ^{2}\geq 4\left( \omega ^{2}/m\right) ,$ $\alpha $ is real and
negative, i. e. for $t\rightarrow \infty ,$ $x$ monotonically goes to zero,
as requiered for a stable system. However, for $\left( \gamma /m\right)
^{2}<4\left( \omega ^{2}/m\right) ,\alpha $ is complex, which means that
approach of $x$ to zero takes place with periodically decreasing amplitude.

Equation (\ref{int01}) describes a pure mechanical system in the classical sense, i.e., zero
temperature, while for quantum description the fluctuations persist even in the zero temperature limit. For non-zero temperature, the deterministic equation (\ref%
{int01}) has to be supplemented by thermal noise $\eta (t),$%
\begin{equation}
m\frac{d^{2}x}{dt^{2}}+\gamma \frac{dx}{dt}+\omega ^{2}x=\eta \left( t\right)
\label{int02}
\end{equation}%
where $\eta \left( t\right) $ is a random variable with zero mean $\langle
\eta \left( t\right) \rangle =0$ and a two-point correlation function $%
\langle \eta (t)\eta (t^{\prime })\rangle =2D\delta (t-t^{\prime })$, which
for thermal noise must satisfy the fluctuation-dissipation theorem \cite{lan}
$\langle \eta ^{2}\left( t\right) \rangle =4\gamma \kappa T,$ where $\kappa $
is the Boltzmann constant. For $m=0$ and $\omega =0,$ Eq.~(\ref{int02}),
 describes an over damped Brownian particle, first introduced by
Einstein more than $100$ years ago.

Another generalization of Eq.~(\ref{int01}) consists in adding external
noise, which enters the equation of motion multiplicatively. For example,
random damping yields 
\begin{equation}
m\frac{d^{2}x}{dt^{2}}+\left[ 1+\xi \left( t\right) \right] \gamma \frac{dx}{%
dt}+\omega ^{2}x=\eta \left( t\right) .  \label{int03}
\end{equation}%
%
This equation was first used for the problem of water waves influenced by a
turbulent wind field \cite{west}. By replacing the coordinate $x$ and time $%
t $ by the order parameter and coordinate, respectively, Eq.~(\ref{int01})
can\ be transformed into the stationary linearized Ginzburg-Landau equation with a convective
term, which describes phase transitions in moving systems \cite{31n}. There
are an increasing number of problems in which particles advected by the mean
flow pass through the region under study. These include problems of phase
transition under shear \cite{32n}, open flows of liquids \cite{33n},
Rayleigh-Benard and Taylor-Couette problems in fluid dynamics \cite{34n},
dendritic growth \cite{35n}, chemical waves \cite{36n}, and the motion of
vortices \cite{30n}.

There is also a different  type of Brownian motion, in which
the surrounding molecules are capable not only of colliding with the
Brownian particle, but also adhere to it for some random time, thereby
changing its mass~\cite{GI}. Such a process is described by the following stochastic
equation 
\begin{equation}
m\left[ 1+\xi \left( t\right) \right] \frac{d^{2}x}{dt^{2}}+\gamma \frac{dx}{%
dt}+\omega ^{2}x=\eta \left( t\right) .  \label{int04}
\end{equation}%
%
There are many situations in chemical and biological solutions in which the
surrounding medium contains molecules which are capable of both colliding
with the Brownian particle and also adhering to it for a random time. There
are also some applications of a variable-mass oscillator \cite{abdalla}.
Modern applications of such a model include a nano-mechanical resonator
which randomly absorbs and desorbs molecules \cite{khasin}. The diffusion of
clusters with randomly growing masses has also been considered \cite{luc}.
There are many other applications of an oscillator with a random mass \cite%
{lam}, including ion-ion reactions \cite{gad}-\cite{gad1}, electrodeposition 
\cite{per}, granular flow \cite{gol}, cosmology \cite{benz}-\cite{weid}, film
deposition \cite{kai}, traffic jams \cite{nag}-\cite{benn}, and the stock
market \cite{aus}-\cite{aus1}.

In this paper we further generalize Eq.~(\ref{int01}) to include the case of
all three previously mentioned sources of noise, the additive part of Eq.~(%
\ref{int02}) and the multiplicative parts of Eqs.~(\ref{int03}-\ref{int04}).
Such an equation will describe a coarse-grained situation when a particle is
affected by random kicks from its nearby environment (additive noise),
adhesion of the molecules in the environment (random mass) and changes in
the nearby environment (random friction). While additive random noise is
usually taken to be a Gaussian $\delta $ correlated (i.e. white) noise, this
is not the case for multiplicative noise. It is natural to include
correlations for the multiplicative part, since for example it can take some
time for the attached molecule to return to the environment. Another
complication is the value of the noise. While the random additive kick can
be of any magnitude and sign (i.e. $\pm $), the multiplicative noise does
not have such luxury. Indeed, for the random mass case, a large negative
value of the noise would imply a non-physical negative mass. Although
friction can attain negative magnitude, it is much more common for friction
to be strictly positive. To overcome such restrictions, we use exponentially
correlated dichotomous noise~for  multiplicative noises \cite{GitBook}. A
noise $\xi (t)$ is called dichotomous when it randomly jumps between two
states and its correlation function $\langle \xi (t^{\prime })\xi (t^{\prime
\prime })\rangle $ decays exponentially. The advantage of such a choice for
the noise is that it is not only correlated and bounded, it is also simple
enough to serve as a test case for more complicated noise~\cite{Bena}.

The paper is structured as follows. In Sec.~\ref{massanddampSec}, we
introduce the generalization of Eq.~(\ref{int01}) for the case of random
mass and random damping. The specific noise and the main mathematical tool
(Shapiro-Loginov formula) are described. Section ~\ref{momentsSec} is
devoted to the calculation of the first and second moments of $x$. For each
moment, two stability criteria are discussed, using the roots of an
appropriate characteristic polynomial. The question of response to an
external time-dependent periodic driving force is addressed in Sec.~\ref%
{response}. We use examples of strictly random mass and strictly random
friction to explain various types of observed stochastic resonances.

\section{Random Mass and Random Damping}

\label{massanddampSec} We start with the generalization of the equation of a
linear damped oscillator as previously described. In our generalization the
noise perturbs both the mass of the oscillator and the friction 
\begin{equation}
m(1+\xi _{1}(t))\frac{d^{2}x}{dt^{2}}+\gamma (1+\xi _{2}(t))\frac{dx}{dt}%
+\omega ^{2}x=\eta (t).  \label{geneq01}
\end{equation}%
%
The additive noise is taken to be zero average, $\delta $ correlated $%
\langle \eta (t_{1})\eta (t_{2})\rangle =2D\delta (t_{1}-t_{2})$ and it is
uncorrelated with the multiplicative noise terms $\langle \eta (t_{1})\xi
_{1}(t_{2})\rangle =\langle \eta (t_{1})\xi _{2}(t_{2})\rangle =0$. The
multiplicative noise terms are both assumed to be symmetrical dichotomous
noise with two-point correlation function 
\begin{equation}
\langle \xi _{1}(t_{1})\xi _{1}(t_{2})\rangle =\sigma _{1}^{2}
\exp (-\lambda _{1}|t_{1}-t_{2}|),\quad \langle \xi
_{2}(t_{1})\xi _{2}(t_{2})\rangle =\sigma _{2}^{2}\exp
(-\lambda _{2}|t_{1}-t_{2}|).  \label{geneq02}
\end{equation}%
%
We further assume that the multiplicative noise terms are uncorrelated $%
\langle \xi _{1}(t_{1})\xi _{2}(t_{2})\rangle =0$. An advantage of treating
the noise as symmetrical dichotomous noise is that it allows one to obtain
results for the case of white noise. In the limit $\lambda _{1}\rightarrow
\infty $ (with constant $\sigma _{1}^{2}/\lambda =D_{1}$), the noise $\xi
_{1}$ transforms to white (i.e. $\delta $) correlated noise (a similar
transformation holds of $\xi _{2}$). Before turning to the calculation of
the moments of $x$, we mention the central tool we apply to obtain a
solution. For an exponentially correlated stochastic process $\xi $ (i.e.
Eq.~(\ref{geneq02})) and some general function of the process $g(\xi )$, the
following relation holds 
\begin{equation}
\left( \frac{d}{dt}+\lambda \right) ^{n}\langle \xi g\rangle =\langle \xi 
\frac{d^{n}g}{dt^{n}}\rangle ,  \label{geneq03}
\end{equation}%
%
where $n$ is a positive integer. Equation~(\ref{geneq03}) is the
Shapiro-Loginov formula~\cite{shapiro} and its generalization for the case of two sources of noise is $\left(
d/dt+(\lambda _{1}+\lambda _{2})\right) ^{n}\langle \xi _{1}\xi _{2}g\rangle
=\langle \xi _{1}\xi _{2}d^{n}g/dt^{n}\rangle $.

\section{ Calculation of the moments}

\label{momentsSec}

\subsection{Behavior of the Mean}

\label{meanSec}

We perform four operations upon Eq.~(\ref{geneq01}) : (i) averaging with
respect to the noise; (ii) multiplying by $\xi _{1}(t)$ and averaging; (iii)
multiplying by $\xi _{2}(t)$ and averaging; (iv) multiplying by $\xi
_{1}(t)\xi _{2}(t)$ and averaging. 
By exploiting the property of dichotomous noise $\xi _{1}(t)\xi
_{1}(t)=\sigma _{1}^{2}$ and $\xi _{2}(t)\xi _{2}(t)=\sigma _{2}^{2}$ 
and applying the Shapiro-Loginov formula (as given by Eq.~(\ref{geneq03}))
we obtain 
\begin{equation}
\mathbf{a}\left( \frac{d}{dt}\right) \cdot 
\begin{pmatrix}
\langle \xi _{1}x\rangle \\ 
\langle \xi _{2}x\rangle \\ 
\langle \xi _{1}\xi _{2}x\rangle \\ 
\langle x\rangle%
\end{pmatrix}%
=0
\label{averresfi01}
\end{equation}%
%
where 
\begin{equation}
\mathbf{a}\left( \frac{d}{dt}\right) =%
\begin{pmatrix}
0 & \left( b_{2}^{2}+\frac{\gamma }{m}b_{2}+\frac{\omega ^{2}}{m}\right) & 
b_{3}^{2} & \sigma _{2}^{2}\frac{\gamma }{m}\frac{d}{dt} \\ 
\left( b_{1}^{2}+\frac{\gamma }{m}b_{1}+\frac{\omega ^{2}}{m}\right) & 0 & 
\frac{\gamma }{m}b_{3} & \sigma _{1}^{2}\frac{d^{2}}{dt^{2}} \\ 
b_{1}^{2} & \frac{\gamma }{m}b_{2} & 0 & \left( \frac{d^{2}}{dt^{2}}+\frac{%
\gamma }{m}\frac{d}{dt}+\frac{\omega ^{2}}{m}\right) \\ 
\sigma _{2}^{2}\frac{\gamma }{m}b_{1} & \sigma _{1}^{2}b_{2}^{2} & \left(
b_{3}^{2}+\frac{\gamma }{m}b_{3}+\frac{\omega ^{2}}{m}\right) & 0%
\end{pmatrix}%
.  \label{averresfi02}
\end{equation}%
%
In Eq.~(\ref{averresfi02}) $b_{1}=(d/dt+\lambda _{1})$, $b_{2}=(d/dt+\lambda
_{2})$ and $b_{3}=(d/dt+(\lambda _{1}+\lambda _{2}))$. The well known
Cramer's rule yields 
\begin{equation}
\left\vert \mathbf{a}\left( \frac{d}{dt}\right) \right\vert \langle x\rangle
=0.  \label{averresfi03}
\end{equation}%
%
Substituting the expressions for $a_{ij}$ yields a differential equation of
eighth order with constant coefficients $\sum_{i=0}^{i=8}c_{8-i}\frac{d^{i}\langle x\rangle }{dt^{i}}=0$.
%

\begin{figure}[t]
\begin{center}
	\begin{subfigure}[b]{0.5\textwidth}
                \includegraphics[width=\textwidth]{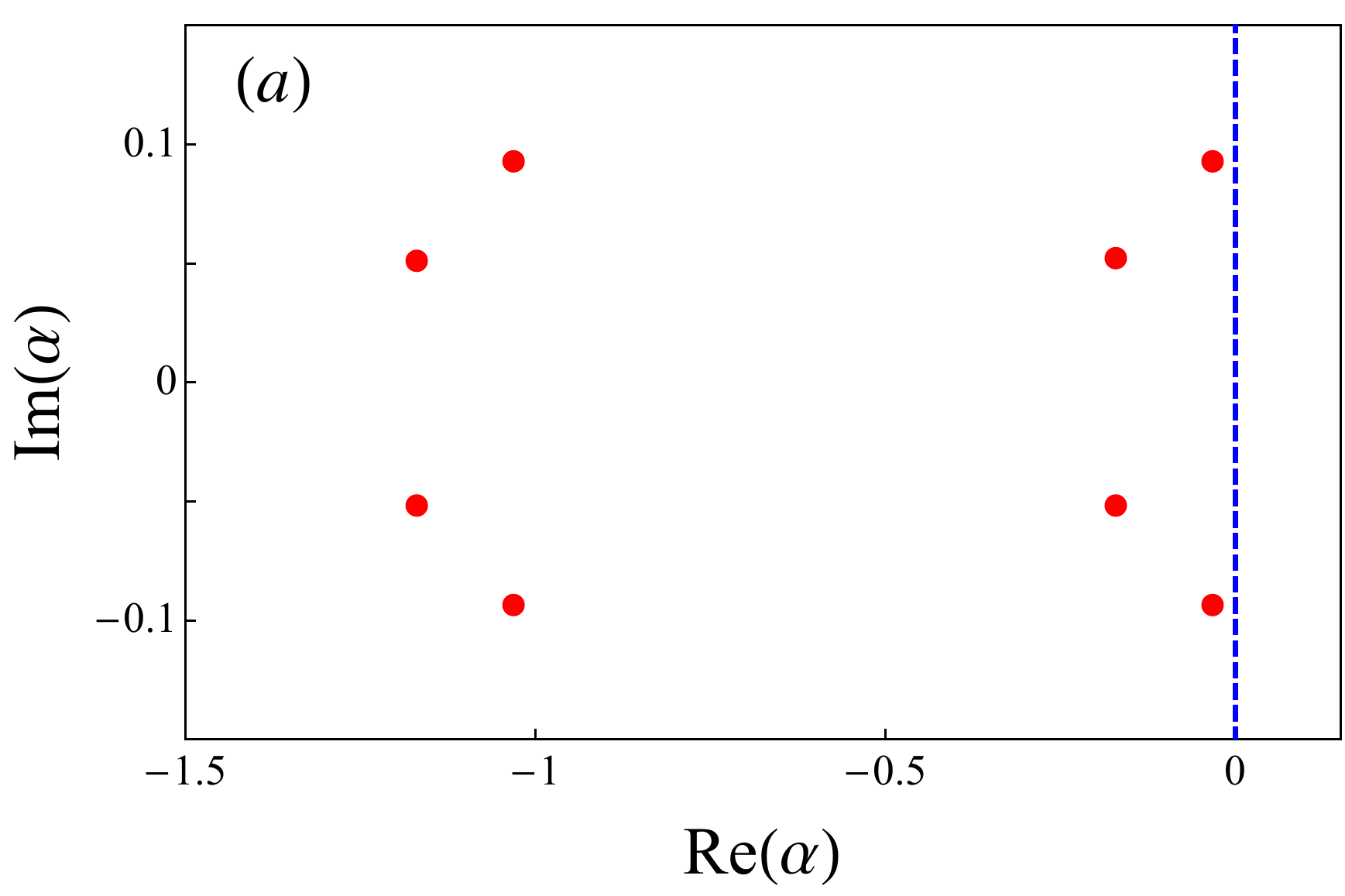}
              
        \end{subfigure}%
        ~
        \begin{subfigure}[b]{0.5\textwidth}
                \includegraphics[width=\textwidth]{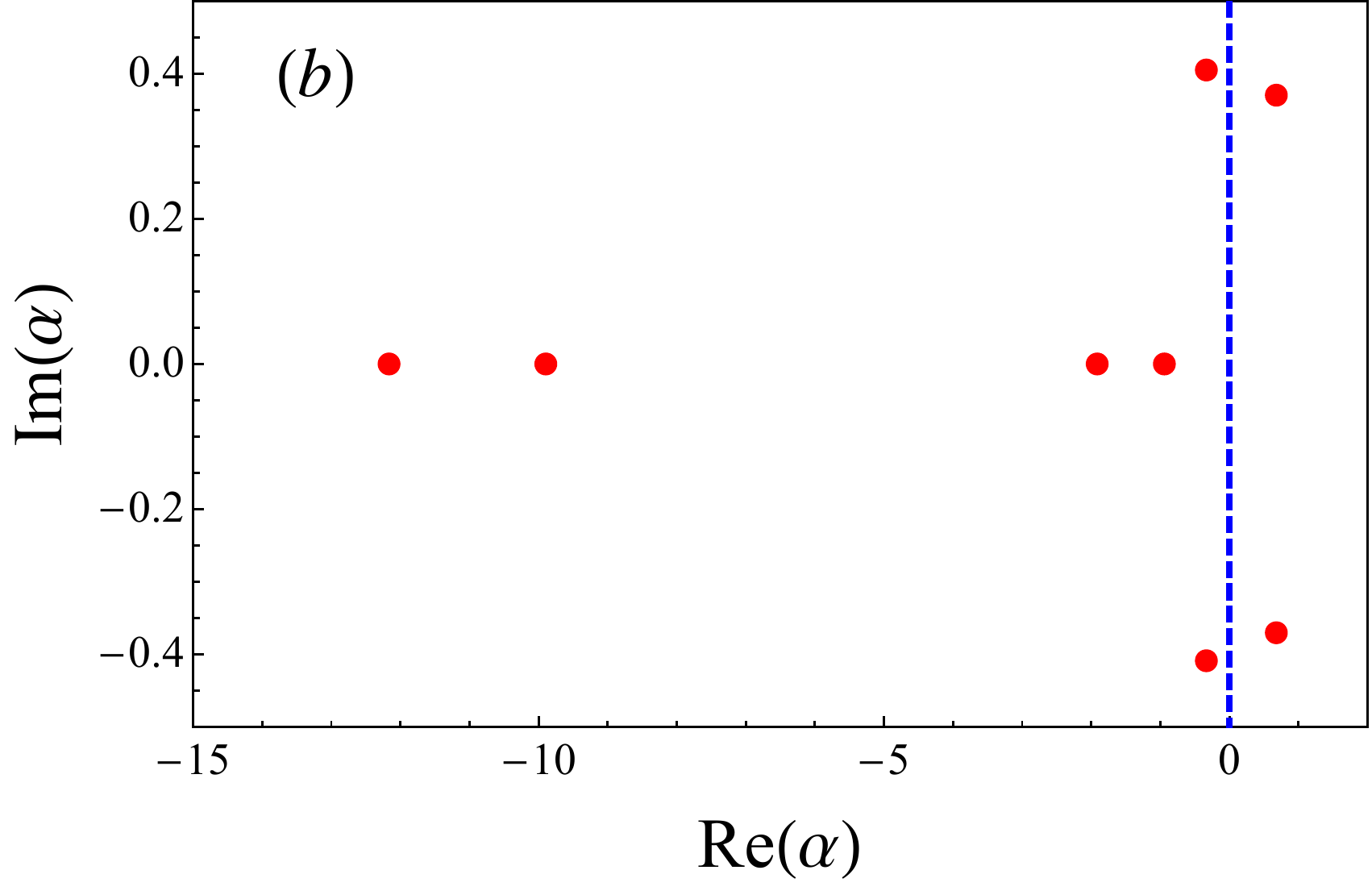}
              
        \end{subfigure}%

\end{center}
\caption{
Values of 
$\alpha ^{\prime }$s which satisfy $\left\vert 
\mathbf{a}\left( \alpha \right) \right\vert =0$, plotted on the complex plane for two different sets of parameters (each dot represent  different $\alpha$) :
{\bf(a)} $\gamma/m=0.1$, $\omega^2/m=0.1$, $\lambda_1=1$, $\lambda_2=0.1$, $\sigma_1=0.1$, $\sigma_2=0.9$.
{\bf(b)} $\gamma/m=4$, $\omega^2/m=1$, $\lambda_1=1$, $\lambda_2=1$, $\sigma_1=0.1$, 
$\sigma_2=1.5$.
The dashed line separates the complex plane into $\text{Re}[\alpha]<0$ and $\text{Re}[\alpha]>0$ .
}
\label{figmean01}
\end{figure}
\begin{figure}[t]
\begin{center}
\includegraphics[width=0.9\textwidth]{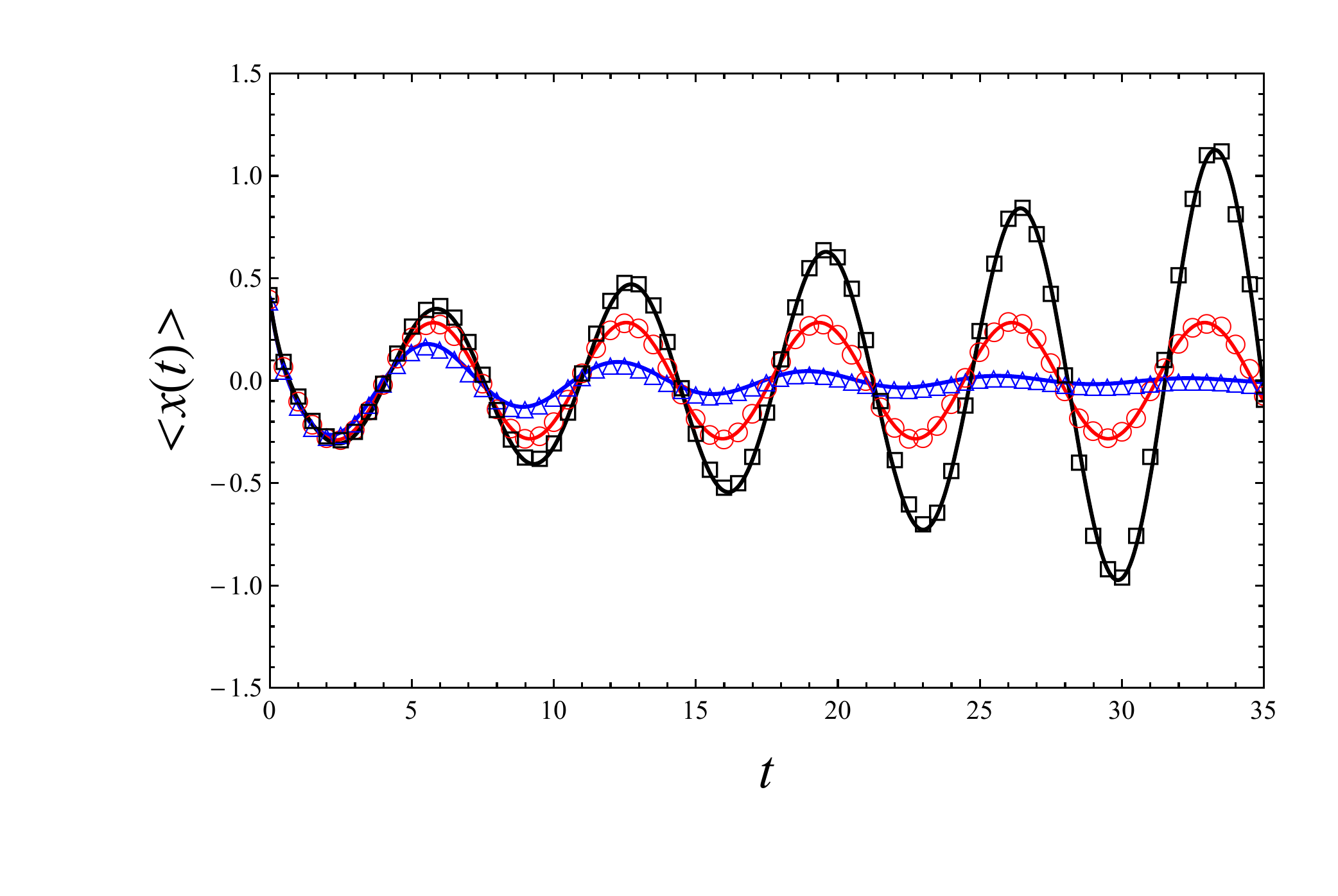}
\end{center}
\caption{
Temporal behavior of $\langle x(t) \rangle$ for three different values of the random damping noise strength $\sigma_2$ while  other parameters are kept constant 
$\gamma /m=1$,$\omega ^{2}/m=1$,$\lambda _{1}=1$,$\lambda_{2}=1$,$\sigma _{1}=1/10$.
The thick lines are the solutions of Eq.~(\ref{averresfi03}) while the symbols are obtained from numerical simulation of the process. Triangles ($\sigma_2=1.45$) are below the transition to instability, circles ($\sigma_2=1.612..$) at the transition and squares 
($\sigma_2=1.7$) above the transition. The numerical data (symbols) was obtained by simulating $10^6$ realizations of the process, each simulation performed by drawing the random times between  switches of $1\pm \sigma_{1}$ from an exponential distribution and similarly drawing random times between the switches of  $1\pm \sigma_{2}$. During the instances when neither of the noises switched, the system was forwarded in time by exact integration.
}
\label{figtrans}
\end{figure}

Seeking a solution of the form $e^{\alpha t},$ we obtain
that $\alpha $ is a solution of $\left\vert \mathbf{a}\left( \alpha \right)
\right\vert =0$ 
The expressions for various $\alpha $ can only be found numerically. The
stability of the system can be explored by studing the asymptotic behavior
of $\langle x\rangle $. The behavior will be stable if $\langle x(t)\rangle
\rightarrow 0$ as $t\rightarrow \infty $. 
%
The general criteria for
stability is the condition that for all $\alpha ,$ which satisfy $\left\vert 
\mathbf{a}\left( \alpha \right) \right\vert =0,$ the value of $\alpha $ has
a negative real part. The Routh-Hurowitz theorem~\cite{Hurwitz} provides the
condition for all the roots of polynomial to have a negative real part. The
condition involves the calculation of the determinants of matrices up to $%
15\times 15$ and is rather cumbersome. Instead, one can plot the various
roots $\alpha $ on the complex plane and investigate their positions for
various values of the parameters $\gamma /m,\omega ^{2}/m,\lambda
_{1},\lambda _{2},\sigma _{1},\sigma _{2}$ . In Fig.~\ref{figmean01}, two
examples are presented. In panel (a) the configuration of the roots is such
that for all eight $\alpha $, $\text{Re}[\alpha ]<0$ and eventually $\langle
x\rangle $ decays to zero. When there is at least one $\alpha $ for which $%
\text{Re}[\alpha ]\geq 0$, i.e. panel (b), $\langle x\rangle $ does not
converge to zero and the behavior is not stable in the mean sense. We note
that the transition to instability can be achieved in various ways. There
are various configurations of parameters for which exactly at the transition
point, $\langle x\rangle $ will exhibit stable oscillations. Specifically,
this occurs for $\gamma /m=1,\omega ^{2}/m=1,\lambda _{1}=1,\lambda
_{2}=1,\sigma _{1}=1/10,\sigma _{2}=1.612443...$. In Fig.~\ref{figtrans} the behavior of 
$\langle x(t) \rangle$ is plotted as function of time for the mentioned parameters and three 
different values of $\sigma_{2}$. Below the transition to instability ($\sigma_{2}=1.45$), 
decaying oscillations occur. At the instability ($\sigma _{2}=1.612...$) the oscillations are stable, and above the transition ($\sigma_{2}=1.7$) the oscillations are diverging. Those results were obtained both by solution of Eq.~(\ref{averresfi03}) and numerical simulation of the stochastic process.

\subsection{Behavior of $\langle x^2 \rangle$}

The stability criteria in the mean sense, as described in the previous
section, can be rather unsatisfying. Indeed, the convergence of the mean to
zero in the long run does not provide any certainty that the process $x$ (as
described by Eq.~(\ref{geneq01})) will be in the vicinity of zero. For
example, the simple random walk starting from zero will on average be at
zero, but the divergence of the second moment of a simple random walk
produces very long excursions towards $\pm \infty $. It is thus preferable to
obtain conditions for stability based on the behavior of the second moment $%
\langle x^{2}\rangle $. 
Generally the divergence of specific moment $\langle x^{n}(t) \rangle$ depends on the properties of the tail of the time dependent distribution of  $x$, $P(x,t)$. The case when $P(x,t)$ decays 
as $|x|^{-1-z}$, with $1<z<2$, produces stable solution for the mean but divergence of the 
second comment. The ability to compute the full distribution $P(x,t)$ is beyond the scope of this study (or any other study to the best of our knowledge) and we therefor proceed to the exploration of the second comment.
We note that in the literature~\cite{Mendez,Mankin} the
instability based on the behavior of the second moment is addressed as an
energetic instability. In order to obtain the various possible behaviors of $%
\langle x^{2}\rangle ,$ we now turn to Eq.~(\ref{geneq01}) similarly to what
was done for $\langle x\rangle .$

We rewrite Eq.~(\ref{geneq01}) in the following form 
\begin{equation}
\begin{split}
\frac{dx}{dt}=& y \\
\frac{dy}{dt}=& -\frac{\gamma }{m}\frac{1+\xi _{2}}{1+\xi _{1}}y-\frac{%
\omega ^{2}}{m}\frac{1}{1+\xi _{1}}x-\frac{m}{1+\xi _{1}}\eta (t),
\end{split}
\label{secmom01}
\end{equation}%
%
and then obtain from Eq.~(\ref{secmom01}) three equations after multiplying
them by $x$ and by $y$ and summing up the mixed terms (i.e $ydx/dt+xdy/dt$) 
\begin{equation}
\begin{split}
\frac{dx^{2}}{dt}=& \,2xy \\
\frac{dy^{2}}{dt}=& -\frac{2\gamma }{m}\frac{1+\xi _{2}}{1+\xi _{1}}y^{2}-%
\frac{2\omega ^{2}}{m}\frac{1}{1+\xi _{1}}xy+\frac{2}{m(1+\xi _{1})}y\eta (t)
\\
\frac{dxy}{dt}=& -\frac{\gamma }{m}\frac{1+\xi _{2}}{1+\xi _{1}}xy+y^{2}-%
\frac{\omega ^{2}}{m}\frac{1}{1+\xi _{1}}x^{2}+\frac{1}{m(1+\xi _{1})}x\eta
(t)
\end{split}%
.  \label{secmom02}
\end{equation}%
%
First average Eq.~(\ref{secmom02}) with respect to $\eta $. Since the
multiplicative noise terms $\xi _{1},\xi _{2}$ are uncorrelated with $\eta $,
we treat them as constants and only need to compute the correlators $\langle
x\eta (t)\rangle _{\eta }$ and $\langle y\eta (t)\rangle _{\eta }$. The
symbol $\langle \dots \rangle _{\eta }$ means average only with respect to $%
\eta $. Since $\eta (t)$ is a Gaussian $\delta $ correlated noise we can
invoke Novikov Theorem~\cite{Novikov} for the correlators. The theorem
states that for a vector $\mathbf{u}=(u_{1},u_{2},\dots ,u_{n})$ of
dimension $n$ and Gaussian $\delta $ correlated noise $\eta (t)$ which
satisfy the following relation 
\begin{equation}
\frac{d\mathbf{u}}{dt}=\mathbf{f}(\mathbf{u})+\mathbf{g}(\mathbf{u})\eta (t),
\label{novikov01}
\end{equation}%
%
where $\mathbf{f}(\mathbf{u})=(f_{1}(\mathbf{u}),f_{2}(\mathbf{u}),\dots
,f_{n}(\mathbf{u}))$ and $\mathbf{g}(\mathbf{u})=(g_{1}(\mathbf{u})g_{2}(%
\mathbf{u}),\dots ,g_{n}(\mathbf{u}))$, the correlators satisfy 
\begin{equation}
\langle g_{i}(\mathbf{u})\eta (t)\rangle _{\eta }=D\sum_{j=1}^{n}\langle 
\frac{\partial g_{i}(\mathbf{u})}{\partial u_{j}}g_{j}(\mathbf{u})\rangle
_{\eta }.  \label{novikov02}
\end{equation}%
%
From Eq.~(\ref{secmom02}), we define $\mathbf{u}=(x^{2},y^{2},xy)$ and $%
\mathbf{g}(\mathbf{u})=(0,\frac{2}{m(1+\xi _{2})}\sqrt{y^{2}},\frac{1}{%
m(1+\xi _{2})}\sqrt{x^{2}}).$ Applying Novikov Theorem yields%
\begin{equation}
\begin{split}
\langle y\eta \rangle _{\eta }=& \frac{D}{m(1+\xi _{1})} \\
\langle x\eta \rangle _{\eta }=& 0 \\
&
\end{split}%
.  \label{novikov03}
\end{equation}%
%
Averaging Eq.~(\ref{secmom02}) with respect to $\eta $ and inserting Eq.~(%
\ref{novikov03}) for the correlators, we obtain 
\begin{equation}
\begin{split}
\frac{d\langle x^{2}\rangle _{\eta }}{dt}-2\langle xy\rangle _{\eta }=& 0 \\
(1+\xi _{1})^{2}\frac{d\langle y^{2}\rangle _{\eta }}{dt}+\frac{2\gamma }{m}%
(1+\xi _{2})(1+\xi _{1})\langle y^{2}\rangle _{\eta }+\frac{2\omega ^{2}}{m}%
(1+\xi _{1})\langle xy\rangle _{\eta }-\frac{2D}{m^{2}}=& 0 \\
(1+\xi _{1})\frac{d\langle xy\rangle _{\eta }}{dt}+\frac{\gamma }{m}(1+\xi
_{2})\langle xy\rangle _{\eta }-(1+\xi _{1})\langle y^{2}\rangle _{\eta }+%
\frac{\omega ^{2}}{m}\langle x^{2}\rangle _{\eta }=& 0 \\
&
\end{split}
\label{secmom03}
\end{equation}%
%
Equation~(\ref{secmom03}) is then treated in the same fashion as Eq.~(\ref%
{geneq01}) in Sec.~\ref{meanSec}. Four operations are performed upon each
line in Eq.~(\ref{secmom03}) : (i) averaging with respect to the noises;
(ii) multiplying by $\xi _{1}(t)$ and averaging; (iii) multiplying by $\xi
_{2}(t)$ and averaging; (iv) multiplying by $\xi _{1}(t)\xi _{2}(t)$ and
averaging. Since all sources of noise are uncorrelated we can switch the
order of averaging. The outcome of the averaging order switching is that we
may treat $\langle x^{2}\rangle _{\eta },\langle y^{2}\rangle _{\eta
},\langle xy\rangle _{\eta }$ as $x^{2},y^{2},xy$ and after applying the
Shapiro-Loginov procedure (Eq.~(\ref{geneq03})), only terms of the type $%
\left( \langle x^{2}\rangle ,\langle y^{2}\rangle ,\langle xy\rangle
,\langle \xi _{1}x^{2}\rangle ,\dots \right) $ remain. The final result of
the averaging is written in matrix form
\begin{equation}
\mathbf{M}\cdot {\vec{X}}={\vec{X}_{0}}
\label{secmom04}
\end{equation}%
%
where $\mathbf{M}$ is given by 
\begin{equation}
\mathbf{M}\left( \frac{d}{dt}\right) =\left( 
\begin{smallmatrix}
\frac{d}{dt} & 0 & -2 & 0 & 0 & 0 & 0 & 0 & 0 & 0 & 0 & 0 \\ 
0 & 0 & 0 & b1 & 0 & -2 & 0 & 0 & 0 & 0 & 0 & 0 \\ 
0 & 0 & 0 & 0 & 0 & 0 & b1 & 0 & -2 & 0 & 0 & 0 \\ 
0 & 0 & 0 & 0 & 0 & 0 & 0 & 0 & 0 & b1 & 0 & -2 \\ 
0 & (1+\sigma _{1}^{2})\frac{d}{dt}+\frac{2\gamma }{m} & \frac{2\omega ^{2}}{%
m} & 0 & 2b_{1}+\frac{2\gamma }{m} & \frac{2\omega ^{2}}{m} & 0 & \frac{%
2\gamma }{m} & 0 & 0 & \frac{2\gamma }{m} & 0 \\ 
0 & 2\sigma _{1}^{2}\frac{d}{dt}+\frac{2\gamma }{m}\sigma _{1}^{2} & \frac{%
2\omega ^{2}}{m}\sigma _{1}^{2} & 0 & (1+\sigma _{1}^{2})b1+\frac{2\gamma }{m%
} & \frac{2\omega ^{2}}{m} & 0 & \frac{2\gamma }{m}\sigma _{1}^{2} & 0 & 0 & 
\frac{2\gamma }{m} & 0 \\ 
0 & \frac{2\gamma }{m}\sigma _{2}^{2} & 0 & 0 & \frac{2\gamma }{m}\sigma
_{2}^{2} & 0 & 0 & (1+\sigma _{1}^{2})b_{2}+\frac{2\gamma }{m} & \frac{%
2\omega ^{2}}{m} & 0 & 2b_{3}+\frac{2\gamma }{m} & \frac{2\omega ^{2}}{m} \\ 
0 & \frac{2\gamma }{m}\sigma _{1}^{2}\sigma _{2}^{2} & 0 & 0 & \frac{2\gamma 
}{m}\sigma _{2}^{2} & 0 & 0 & 2\sigma _{1}^{2}b_{2}+\frac{2\gamma }{m}\sigma
_{1}^{2} & \frac{2\omega ^{2}}{m}\sigma _{1}^{2} & 0 & (1+\sigma
_{1}^{2})b_{3}+\frac{2\gamma }{m} & \frac{2\omega ^{2}}{m} \\ 
\frac{\omega ^{2}}{m} & -1 & \frac{d}{dt}+\frac{\gamma }{m} & 0 & -1 & b_{1}
& 0 & 0 & \frac{\gamma }{m} & 0 & 0 & 0 \\ 
0 & -\sigma _{1}^{2} & \sigma _{1}^{2}\frac{d}{dt} & \frac{\omega ^{2}}{m} & 
-1 & b_{1}+\frac{\gamma }{m} & 0 & 0 & 0 & 0 & 0 & \frac{\gamma }{m} \\ 
0 & 0 & \frac{\gamma }{m}\sigma _{2}^{2} & 0 & 0 & 0 & \frac{\omega ^{2}}{m}
& -1 & b_{2}+\frac{\gamma }{m} & 0 & -1 & b_{3} \\ 
0 & 0 & 0 & 0 & 0 & \frac{\gamma }{m}\sigma _{2}^{2} & 0 & -\sigma _{1}^{2}
& \sigma _{1}^{2}b_{2} & \frac{\omega ^{2}}{m} & -1 & b_{3}+\frac{\gamma }{m}
\\ 
&  &  &  &  &  &  &  &  &  &  & 
\end{smallmatrix}%
\right)  \label{secmom05}
\end{equation}%
%
where ${\vec{X}}=\left( \langle x^{2}\rangle ,\langle y^{2}\rangle ,\langle
xy\rangle ,\langle \xi _{1}x^{2}\rangle ,\langle \xi _{1}y^{2}\rangle
,\langle \xi _{1}xy\rangle ,\langle \xi _{2}x^{2}\rangle ,\langle \xi
_{2}y^{2}\rangle ,\langle \xi _{2}xy\rangle ,\langle \xi _{1}\xi
_{2}x^{2}\rangle ,\langle \xi _{1}\xi _{2}y^{2}\rangle ,\langle \xi _{1}\xi
_{2}xy\rangle \right) $ and ${\vec{X_{0}}}=\left(
0,0,0,0,2D/m^{2},0,0,0,0,0,0,0\right) $. Cramer's rule implies 
\begin{equation}
\left\vert \mathbf{M}\left( \frac{d}{dt}\right) \right\vert \langle
x^{2}\rangle =\left\vert \mathbf{M}_{1,5}\left( \frac{d}{dt}\right)
\right\vert \frac{2D}{m^{2}},  \label{secmom06}
\end{equation}%
%
where $|\mathbf{M}_{1,5}|$ is the $\{1,5\}$ minor of matrix $\mathbf{M}$,
i.e determinant of matrix $\mathbf{M}$ where the first column and fifth row
were removed from the matrix. The determinants on both sides of Eq.~(\ref%
{secmom06}) are differential operators and since $\left\vert \mathbf{M}%
_{1,5}\left( d/dt\right) \right\vert $ operates on a constant it can be
replaced by $\left\vert \mathbf{M}_{1,5}\left( 0\right) \right\vert $. The
stable solution is 
\begin{equation}
\langle x_{s}^{2}\rangle =\frac{\left\vert \mathbf{M}_{1,5}\left( 0\right)
\right\vert }{\left\vert \mathbf{M}\left( 0\right) \right\vert }\left(
2D/m^{2}\right) .  \label{secmom07}
\end{equation}%
%
From Eq.~(\ref{secmom07}), it is clear that when $\left\vert \mathbf{M}%
\left( 0\right) \right\vert =0$, the system is not stable and the second
moment diverges. As was the case for $\langle x\rangle $, we can write a
more general condition. 
%
We search a solution of $\left\vert \mathbf{M}\left( \frac{d}{dt}\right) \right\vert \langle x^{2}\rangle =0$ (i.e. the homogeneous part of Eq.~(\ref{secmom07})) in the form of $\exp (\alpha
t). $ This solution will be stable if  $\forall \alpha $ (such that $%
\left\vert \mathbf{M}\left( \alpha \right) \right\vert =0$) $\text{Re}%
[\alpha ]<0$. Then this is the stability criterion and it includes the
special case of $\alpha =0$ that zeros $\left\vert \mathbf{M}\right\vert $.
The search for the criteria of a negative real part of $\left\vert \mathbf{M}%
\left( \alpha \right) \right\vert =0$ can be performed by plotting different
values $\alpha $ on the complex plane and searching for  situations where $\text{Re}[\alpha ]\geq 0$. Specifically for the mentioned case when $\langle x \rangle$ is stable ($\gamma /m=1,\omega ^{2}/m=1,\lambda _{1}=1,\lambda
_{2}=1,\sigma _{1}=1/10,\sigma _{2}=1.45$) the second moment $\langle x^2 \rangle$ will diverge. 

\section{ Response to external driving term.}

\label{response}

We would like to address the question of a response of a noisy oscillator
with random mass and random damping to an external time-dependent driving
term. The external driving term is taken to be a simple sinusoidal form $%
A_{0}\cos \left( \Omega t\right) $. Our general Equation~(\ref{geneq01})
then becomes 
\begin{equation}
m(1+\xi _{1}(t))\frac{d^{2}x}{dt^{2}}+\gamma (1+\xi _{2}(t))\frac{dx}{dt}%
+\omega ^{2}x=\eta (t)+A_{0}\cos \left( \Omega t\right) .  \label{resp01}
\end{equation}%
%
Repeating the steps of Sec.~\ref{meanSec} and using the fact that $A_{0}\cos
\left( \Omega t\right) $ and the multiplicative sources of noise are
uncorrelated , i.e. $\langle \xi _{1}(t)\cos \left( \Omega t\right) \rangle
=\langle \xi _{2}(t)\cos \left( \Omega t\right) \rangle =\langle \xi
_{1}(t)\xi _{2}(t)\cos \left( \Omega t\right) \rangle=0$, we obtain 
\begin{equation}
\mathbf{a}\left( \frac{d}{dt}\right) \cdot 
\begin{pmatrix}
\langle \xi _{1}x\rangle \\ 
\langle \xi _{2}x\rangle \\ 
\langle \xi _{1}\xi _{2}x\rangle \\ 
\langle x\rangle%
\end{pmatrix}%
=%
\begin{pmatrix}
0 \\ 
0 \\ 
A_{0}\cos \left( \Omega t\right) \\ 
0%
\end{pmatrix}%
\label{resp02}
\end{equation}%
%
where $\mathbf{a}\left( d/dt\right) $ is defined by Eq.~(\ref{averresfi02}).
The behavior of $\langle x\rangle $ is given by Cramer's rule 
\begin{equation}
\left\vert \mathbf{a}\left( \frac{d}{dt}\right) \right\vert \langle x\rangle
=-\left\vert \mathbf{a}_{4,3}\left( \frac{d}{dt}\right) \right\vert
A_{0}\cos \left( \Omega t\right) ,  \label{resp03}
\end{equation}%
%
where $\left\vert \mathbf{a}_{4,3}\left( d/dt\right) \right\vert $ is the $%
\{4,3\}$ minor of $\mathbf{a}\left( d/dt\right) $. 
In the limit $t\rightarrow \infty $, when a stable solution for 
$\left\vert \mathbf{a}\left( \frac{d}{dt}\right) \right\vert \langle x\rangle
=0 $ exists and equals to $0$, $\langle x\rangle $ is
given by 
\begin{equation}
\langle x\rangle =A\cos (\Omega t+\phi )  \label{resp06}
\end{equation}%
%
with 
\begin{equation}
A/A_{0}=\sqrt{\frac{\left\vert \mathbf{a}_{4,3}\left( -i\Omega \right)
\right\vert \left\vert \mathbf{a}_{4,3}\left( i\Omega \right) \right\vert }{%
\left\vert \mathbf{a}\left( -i\Omega \right) \right\vert \left\vert \mathbf{a%
}\left( i\Omega \right) \right\vert }}  \label{resp07}
\end{equation}%
%
and 
\begin{equation}
\tan (\phi )=\frac{\left\vert \mathbf{a}_{4,3}\left( -i\Omega \right)
\right\vert \left\vert \mathbf{a}\left( i\Omega \right) \right\vert
+\left\vert \mathbf{a}_{4,3}\left( i\Omega \right) \right\vert \left\vert 
\mathbf{a}\left( -i\Omega \right) \right\vert }{\left\vert \mathbf{a}%
_{4,3}\left( -i\Omega \right) \right\vert \left\vert \mathbf{a}\left(
i\Omega \right) \right\vert -\left\vert \mathbf{a}_{4,3}\left( i\Omega
\right) \right\vert \left\vert \mathbf{a}\left( -i\Omega \right) \right\vert 
}i  \label{resp08}
\end{equation}%
%
The response of $\langle x\rangle $ to the external driving term equals to $%
A/A_{0}$ (Eq.~(\ref{resp07})) when a stable solution exists.

\subsection{Various Aspects of Response}

The expression for the response $A/A_{0}$ depends on seven parameters of the
system and $\Omega $. In order to obtain insight into
the various possible types of behavior, we first treat the two simpler cases
where only one source of multiplicative noises is present, i.e. (i) random
damping (Eq.~(\ref{int04})) or (ii) random mass (Eq.~(\ref{int03})). The
equation describing the case of a random mass and random damping, i.e. Eq.~(%
\ref{geneq01}), reduces to case (i) by taking $\sigma _{2}$ and $\lambda _{2}
$ to zero and to case (ii) by taking $\sigma _{1}$ and $\lambda _{1}$ to
zero. Therefore,  the response to an external periodic driving term for both
simpler cases is provided by $A/A_{0}$ in Eq.~(\ref{resp07}) by setting the
appropriate parameters to zero. We note that both of these simpler cases were
previously treated \cite{GitBook}. In the following mainly the behavior of  $A/A_{0}$ as a function of $\Omega$ is presented. The behavior of $A/A_{0}$ as a function of $\sigma_1$ and $\sigma_2$ is presented in the Appendix.

\begin{figure}[t]
\begin{center}
\begin{subfigure}[b]{0.3255\textwidth}
                \includegraphics[width=\textwidth]{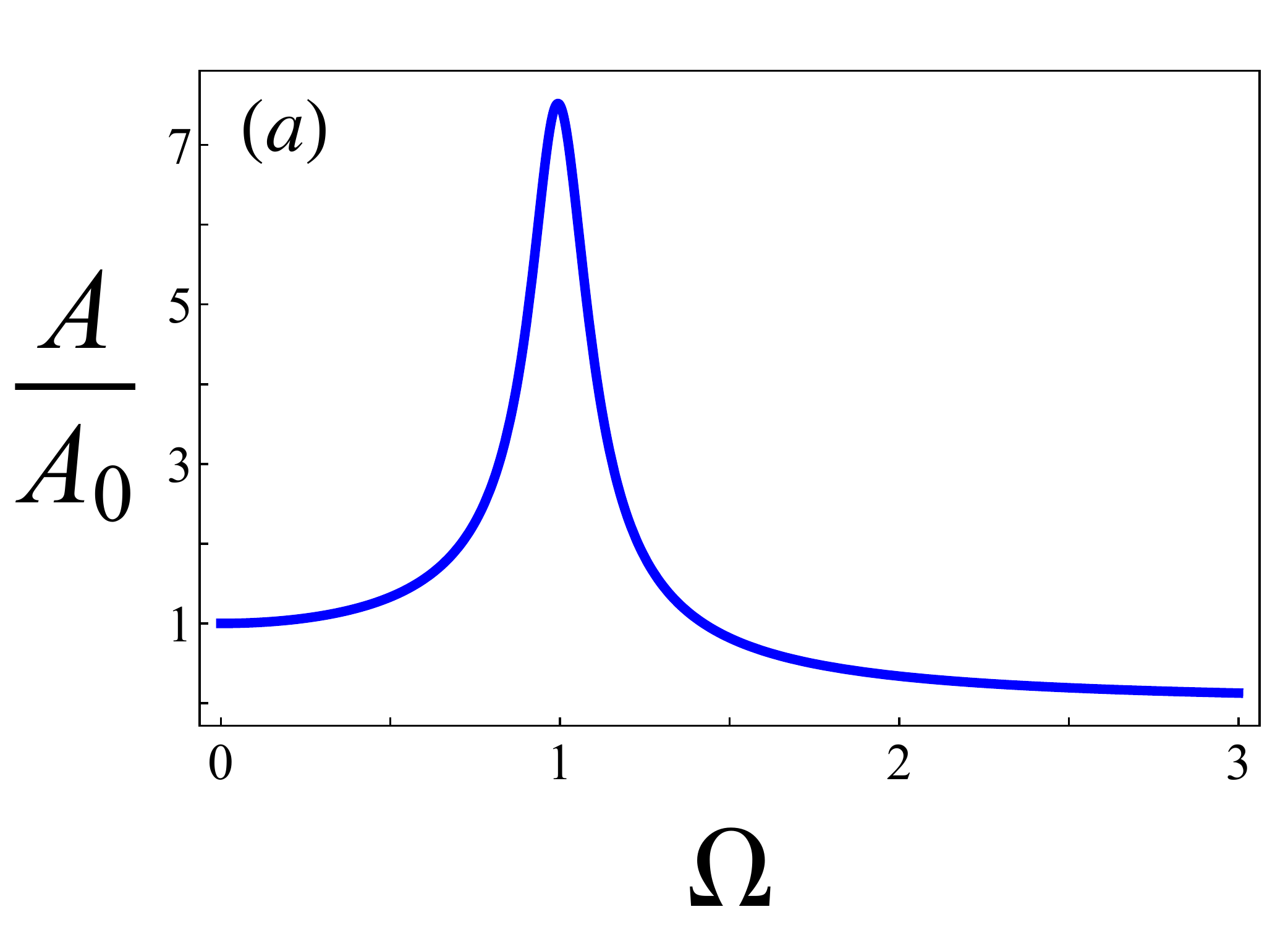}
              
        \end{subfigure} ~ 
\begin{subfigure}[b]{0.3055\textwidth}
                \includegraphics[width=\textwidth]{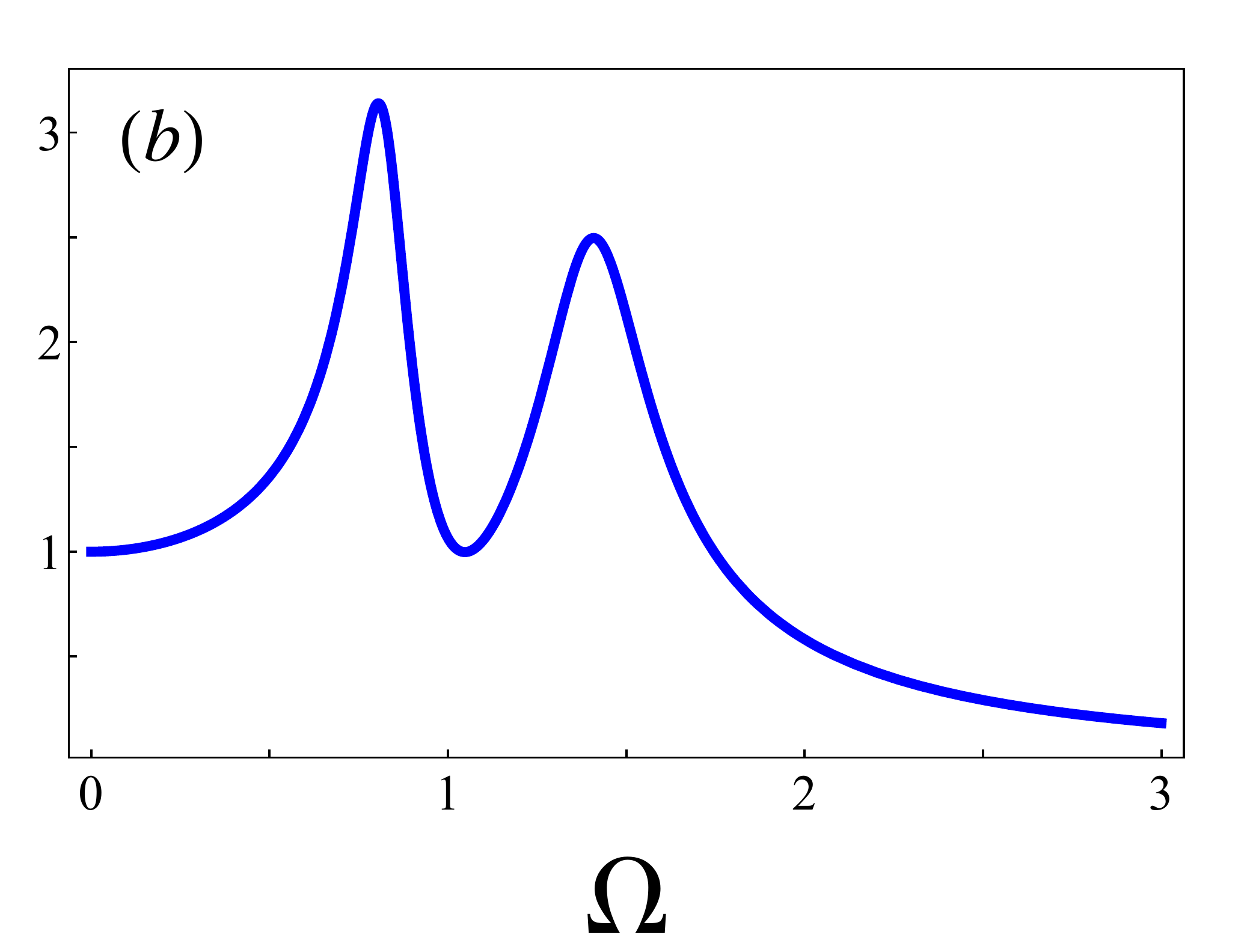}
              
        \end{subfigure} ~%
\begin{subfigure}[b]{0.3055\textwidth}
                \includegraphics[width=\textwidth]{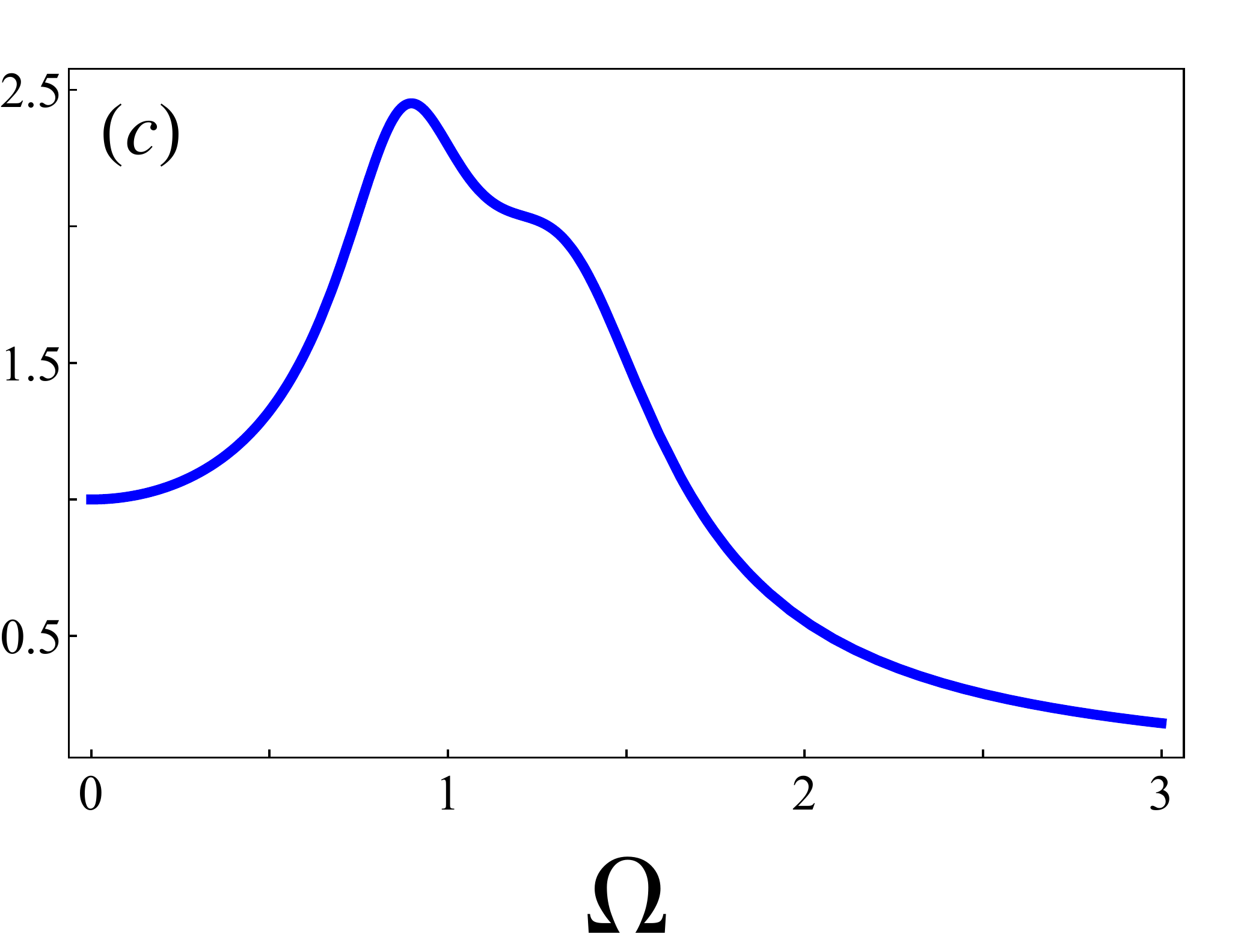}
              
        \end{subfigure}
\end{center}
\caption{ Response $A/A_0$ as a function of angular frequency ($\Omega$) of
the periodic external driving force as given in Eq.~(\ref{resp07}) for
the case of only a random mass. Maximum of $A/A_0$ for specific parameters
of the system describes a resonance between the behavior of $x$ and the
external driving force. \textbf{(a)} $\gamma/m=0.1$, $\omega^2/m=1$, 
$\lambda_1=0.1$,  $\sigma_1=0.1$, $%
\sigma_2=0$. \textbf{(b)} The same parameters as in \textbf{(a)}
except that $\sigma_1=0.5$. \textbf{(c)} The same parameters as in 
\textbf{(b)}, except that $\lambda_1=0.35$. }
\label{figresp01}
\end{figure}

\subsubsection{ Random Mass}
The response for the case of a random mass is presented in Fig.~\ref%
{figresp01}, panels \textbf{(a)-(c)}. In panel \textbf{(a)} a resonance is
found for quite small values of noise strength ($\sigma _{1}^{2}=0.01$).
Increasing the noise strength while keeping the correlation parameter $%
\lambda _{1}$ constant produces an additional maximum for $A/A_{0}$, as
shown in panel \textbf{(b)}. This second resonance is due to the splitting
of the first peak and decreasing its height. Such splitting occurs while the
value of $\lambda _{1}$ is quite small, i.e. large correlation times of $\xi
_{1}$. 

In order to understand the observed effect we notice the fact that random noise $\xi _{1}$ produces two mass values and creates two intrinsic states for the oscillator. In each of the states the oscillator behaves as a simple oscillator with additive noise.  Existence of a resonance will depend on specific parameters of the state : $m_i$, $\gamma_i$ and $\omega$ (subscript $i$ runs over possible state indexes). The resonant frequency $\Omega_R$ (if exisits) is provided by the well known formula~\cite{lanM}
\begin{equation}
\Omega_R = \sqrt{\frac{\omega^2}{m_i}-\frac{\gamma_i^2}{2 m_i^2}}.
\label{simplestate}
\end{equation}
In the case of random mass  $m_1\neq m_2$ and $\gamma_1 = \gamma_2$.
If the oscillator can attain a resonance in both of the states, and the frequencies of those resonances are sufficiently distinct, we expect to observe two resonant frequencies as described in Fig.~\ref{figresp01}. Each of the resonant frequencies will correspond to an intrinsic regime/state of the oscillator and the splitting effect artificially resembles splitting of states in quantum system.
Existence of two states for the oscillator is not sufficient for appearance of two resonant frequencies, the oscillator must also spend a sufficient amount of time (on average) in each of these states in order to attain a resonance. Since the oscillator constantly jumping from one state to the other, the time to build up a "proper" response to external field might be insufficient. The oscillator will jump to the other state where a different response will start to build up. It is thus important that the noise correlation time will be long enough. 
Indeed, this
effect is shown in panel \textbf{(c)} of Fig.~\ref{figresp01} . While
keeping the strength of the noise the same as in \textbf{(b)}, $\lambda _{1}$
was increased and the collapse of the two resonances was obtained.
The case of a random mass can thus contribute to the existence of a single
stochastic resonance, but can also split a single resonance into two
resonances (when the correlation time of the noise is sufficiently
long). Appearance of multiple resonances was also observed in different noisy representations of the 
harmonic oscillator~\cite{Mankin,Peleg}

\begin{figure}[t]
\begin{center}
\begin{subfigure}[b]{0.3355\textwidth}
                \includegraphics[width=\textwidth]{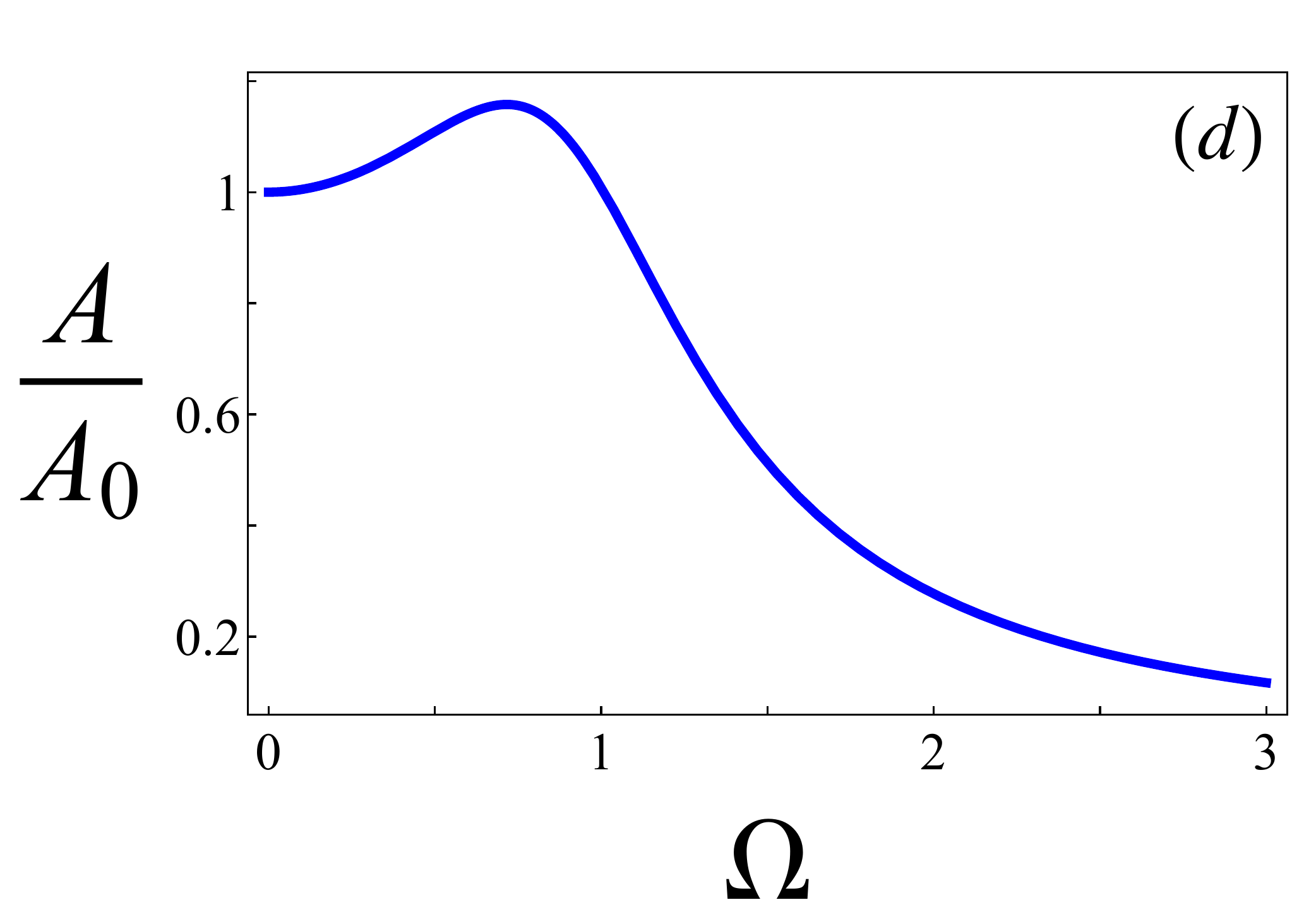}
              
        \end{subfigure} ~ 
\begin{subfigure}[b]{0.3055\textwidth}
                \includegraphics[width=\textwidth]{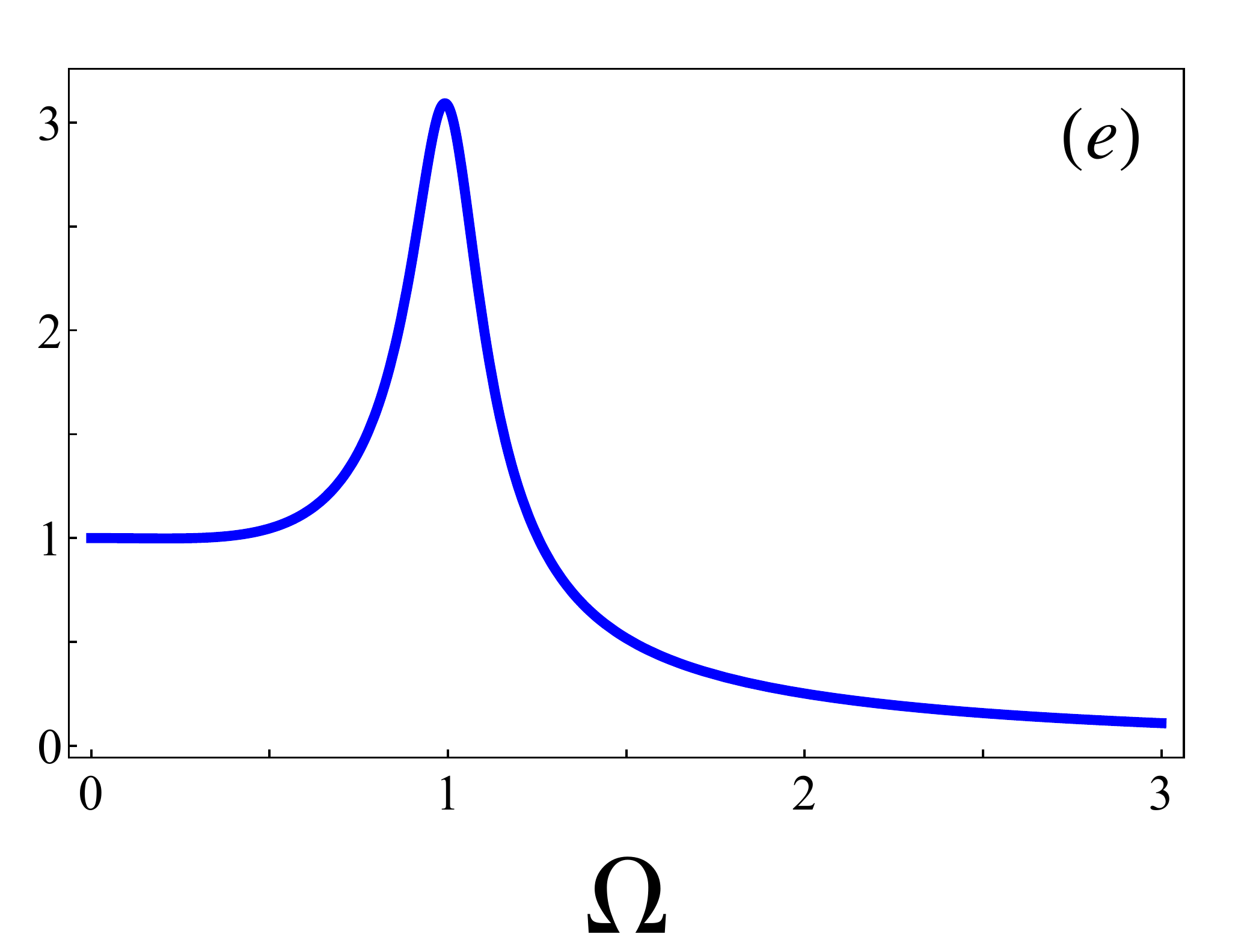}
              
        \end{subfigure} ~ 
\begin{subfigure}[b]{0.3055\textwidth}
                \includegraphics[width=\textwidth]{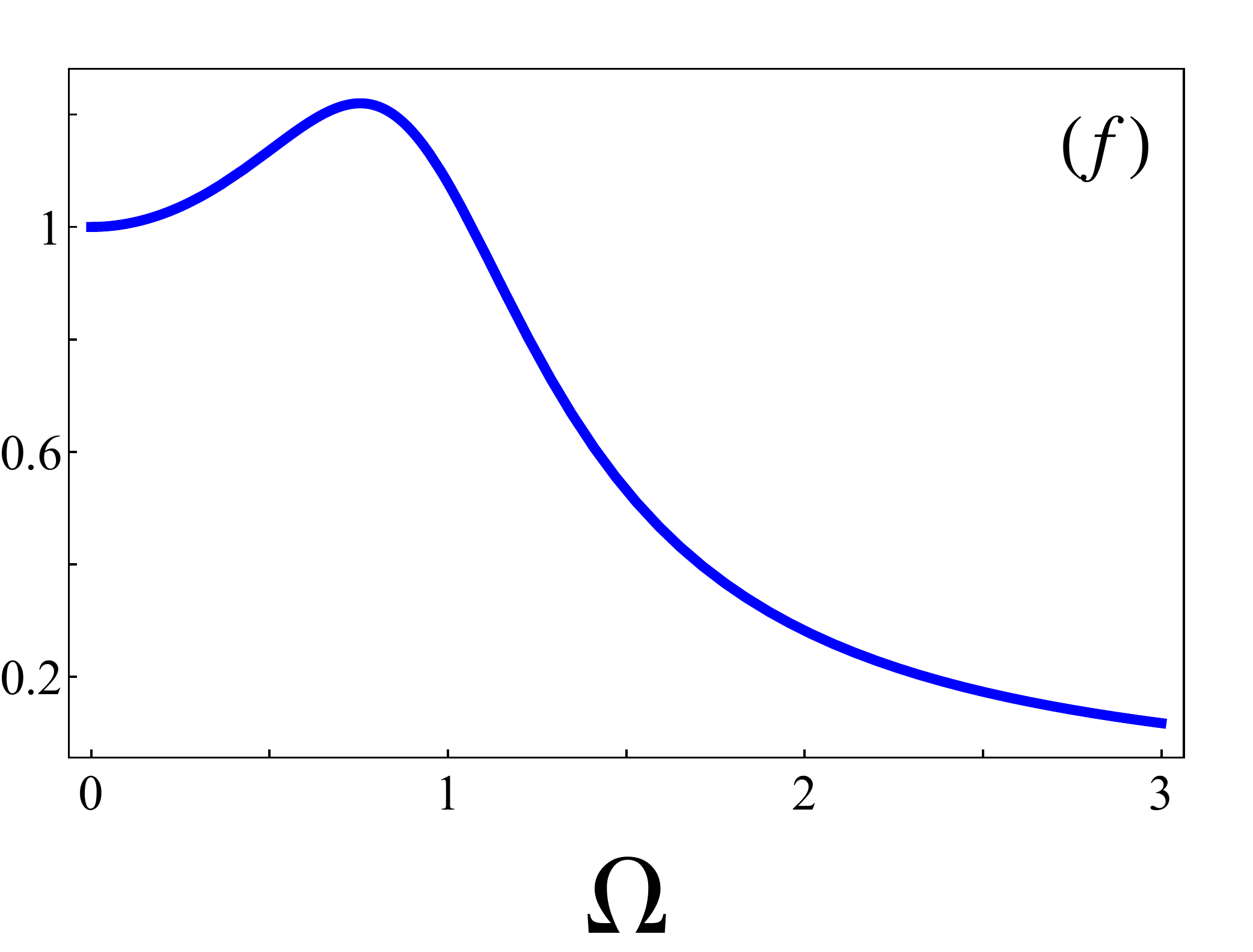}    
        \end{subfigure}
\end{center}
\caption{ Response $A/A_0$ as a function of angular frequency ($\Omega$) of
the periodic external driving as given in Eq.~(\ref{resp07}) for
the case of only random damping. Maximum of $A/A_0$ for specific
parameters of the system describes a resonance between the behavior of $x$
and the external driving force. \textbf{(d)} $\gamma/m=0.1$, $\omega%
^2/m=1$, $\lambda_2=0.1$, $\sigma_1=0$%
, $\sigma_2=0.1$. \textbf{(e)} The same parameters as in \textbf{(d)}%
, except that $\sigma_2=0.9$. \textbf{(f)} The same parameters as in 
\textbf{(e)}, except that $\lambda_2=10$. }
\label{figresp02}
\end{figure}

\subsubsection{ Random Damping}
The response for the case when only random damping exists is presented in
panels (\textbf{(d)-(f)}) of Fig.~\ref{figresp02}. Panel \textbf{(d)} shows
a resonance for a small strength of the multiplicative noise $\xi _{2}$, $%
\sigma _{2}=0.01$. 
In panel \textbf{(e}), the value of $\sigma _{2}^{2}$
was taken to be $0.9,$ yielding a threefold increase in the peak value of $%
A/A_{0}$. 
The effect of resonant frequency splitting, similar to the random mass case, is not observed. 
The oscillator attains two intrinsic states with $\gamma_1\neq\gamma_2$ and $m_1=m_2$.  The functional form of Eq.~(\ref{simplestate}) allows two different resonant frequencies for two states with specific values of $\omega$ and damping. 
But in contrast to the random mass case the difference between two resonant frequencies is not sufficient ($0<\sigma_2<1$). 
Random transitions between two states and the differences in response for each intrinsic state (i.e. decrease in response of one state while increase of the other) will smear presences of two maxima if the maxima frequencies are not sufficiently separated. It seems that for random damping the frequency separation is not sufficient and no splitting is observed.
The increase in the resonance strength due to increase in the damping noise can be explained as a pronounced resonance in a state where the damping is very low (i.e. $\gamma (1-\sigma _{2})$). This response increase is expected to disappear when the time the oscillator spends in a given state will decrease, as explained for the random mass case. Indeed when we decrease this time by increasing $\lambda_2$ the effect disappears. Panel \textbf{(f)} of Fig.~%
\ref{figresp02} shows the disappearance of the threefold increase of the
peak value of the resonance after a significant decrease in the damping
noise correlation time, $\lambda _{2}\rightarrow 10$.

\begin{figure}[t]
\begin{center}
\begin{subfigure}[b]{0.3255\textwidth}
                \includegraphics[width=\textwidth]{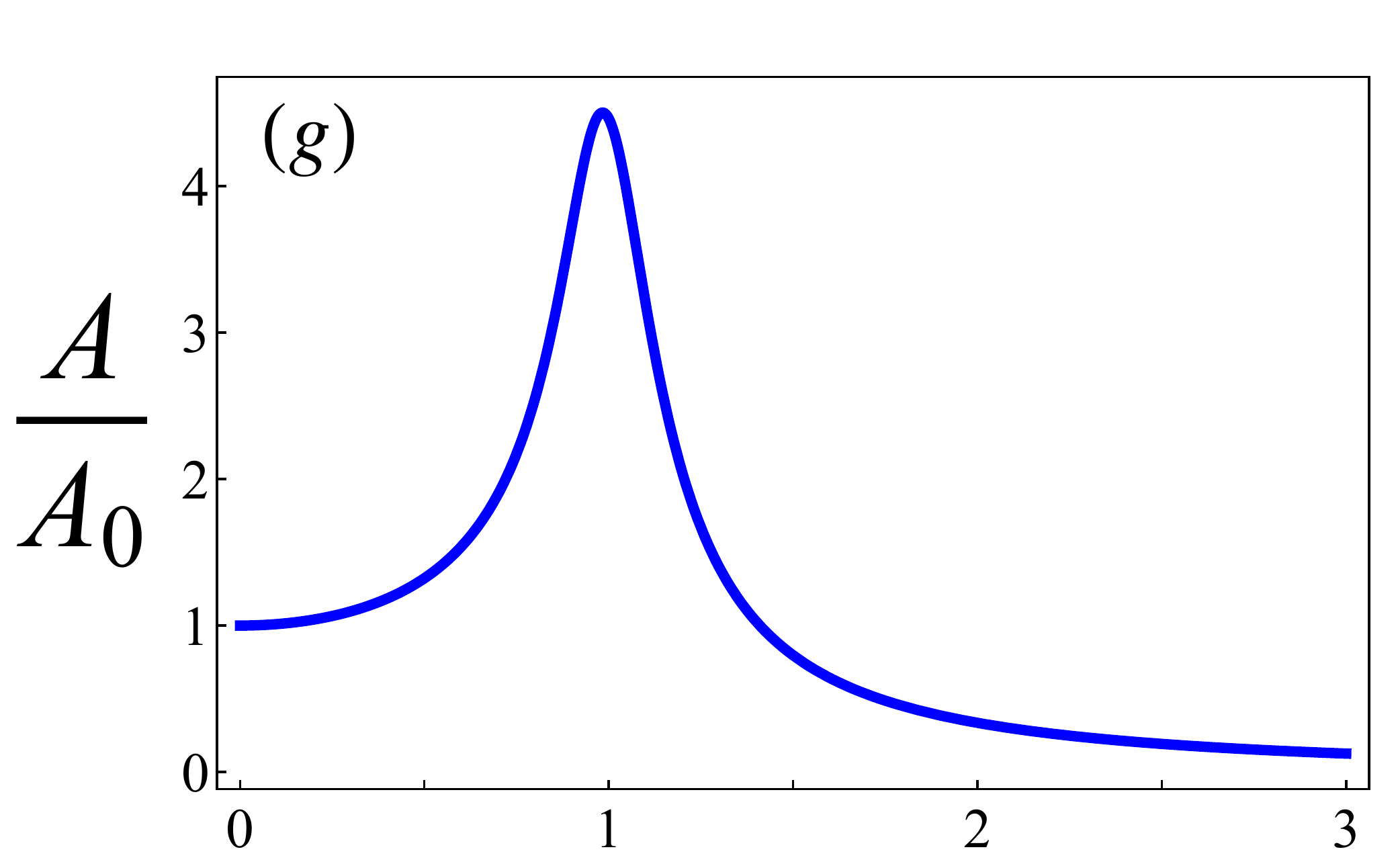}
              
        \end{subfigure} ~ 
\begin{subfigure}[b]{0.3\textwidth}
                \includegraphics[width=\textwidth]{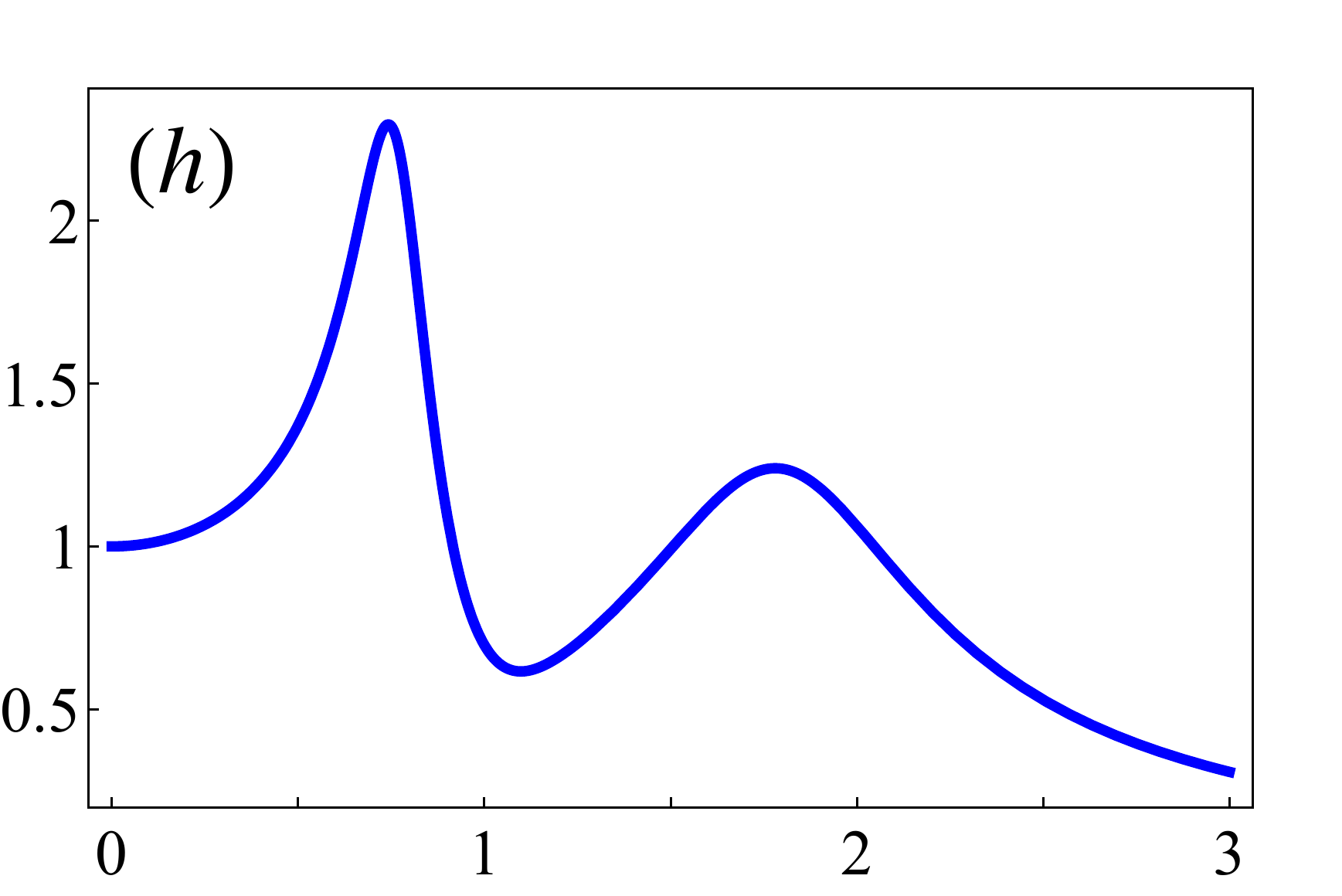}
              
        \end{subfigure} ~ 
\begin{subfigure}[b]{0.3\textwidth}
                \includegraphics[width=\textwidth]{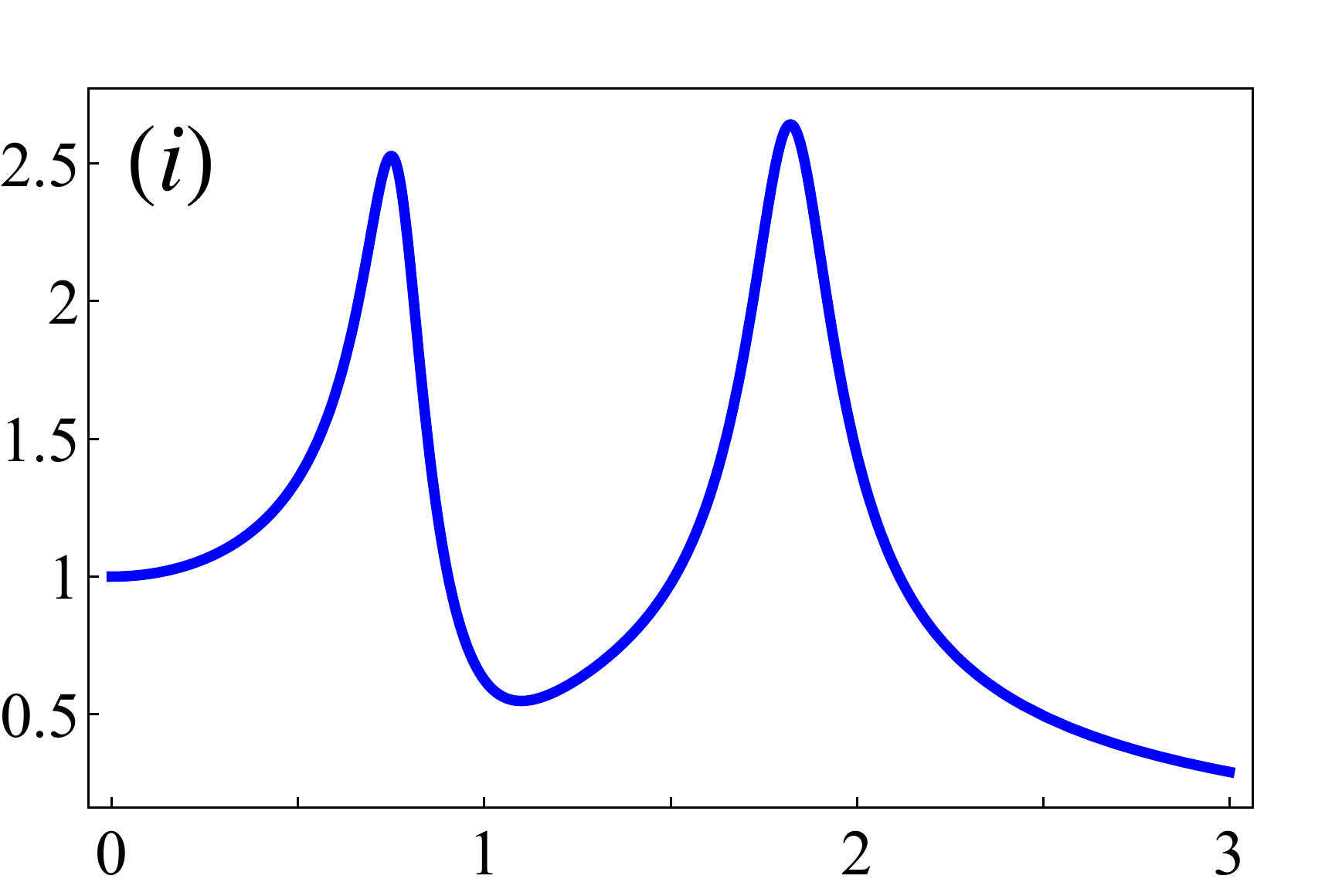}    
        \end{subfigure} \\[0pt]
\begin{subfigure}[b]{0.3255\textwidth}
                \includegraphics[width=\textwidth]{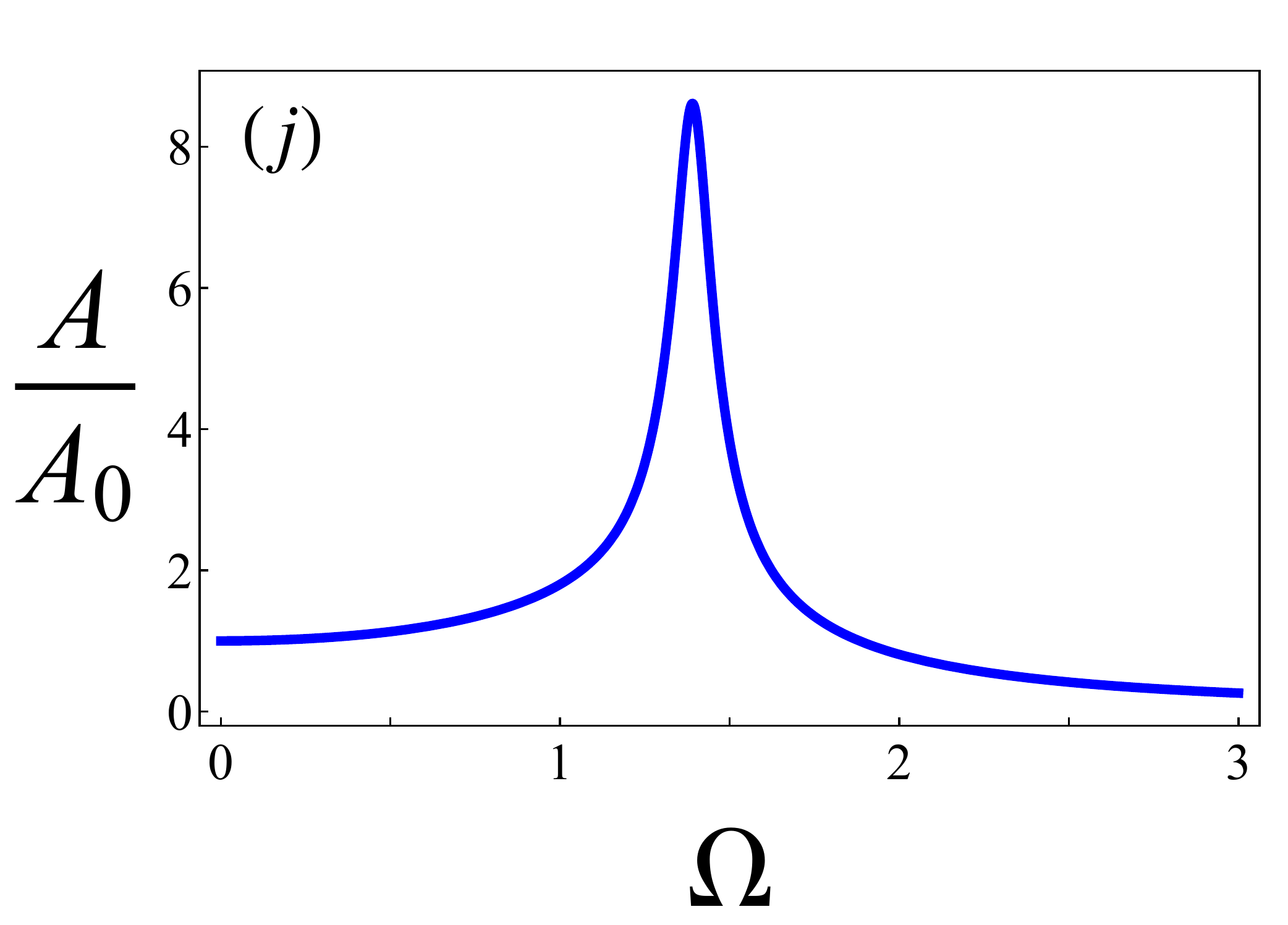}
              
        \end{subfigure} ~ 
\begin{subfigure}[b]{0.298\textwidth}
                \includegraphics[width=\textwidth]{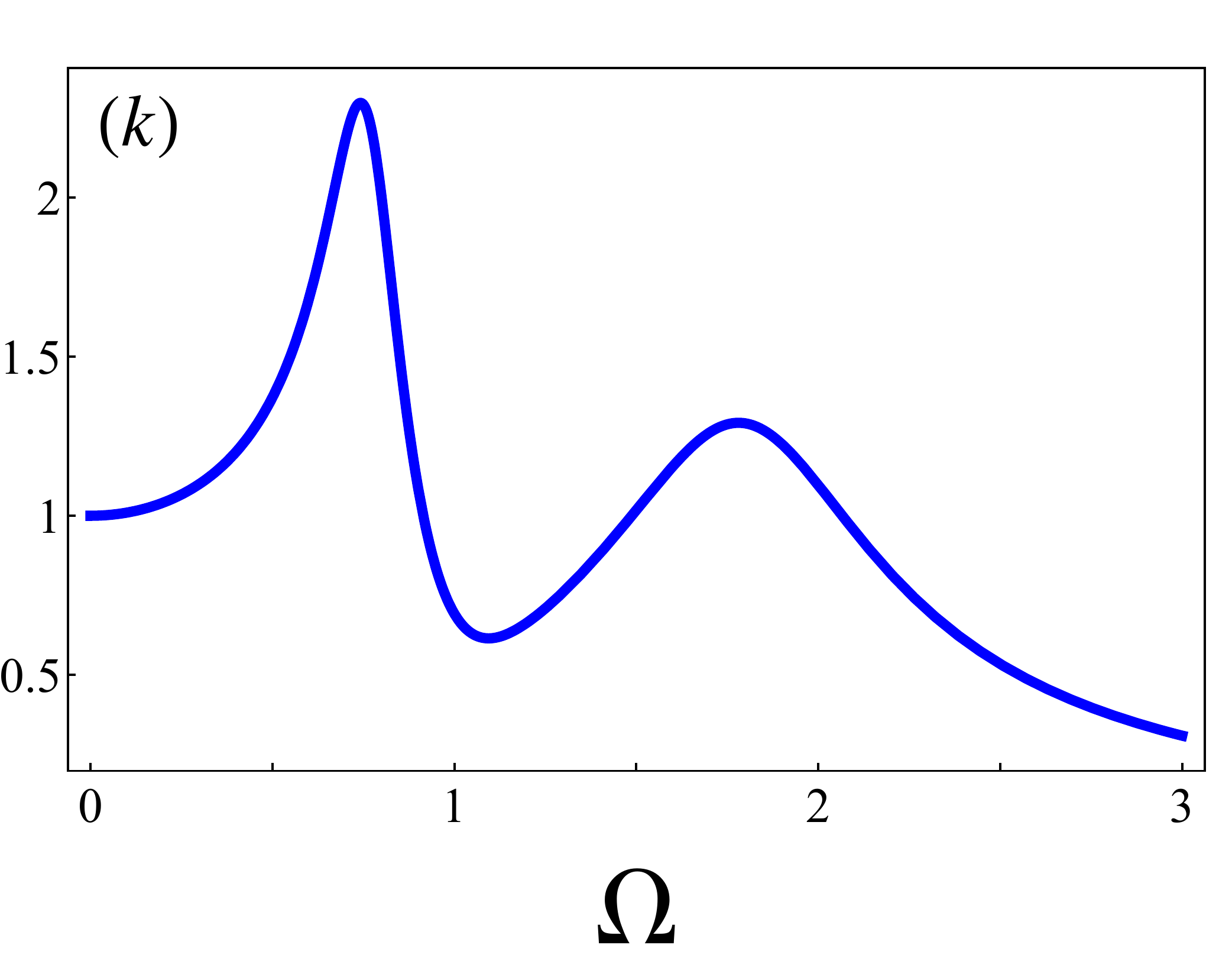}
              
        \end{subfigure} ~ 
\begin{subfigure}[b]{0.306\textwidth}
                \includegraphics[width=\textwidth]{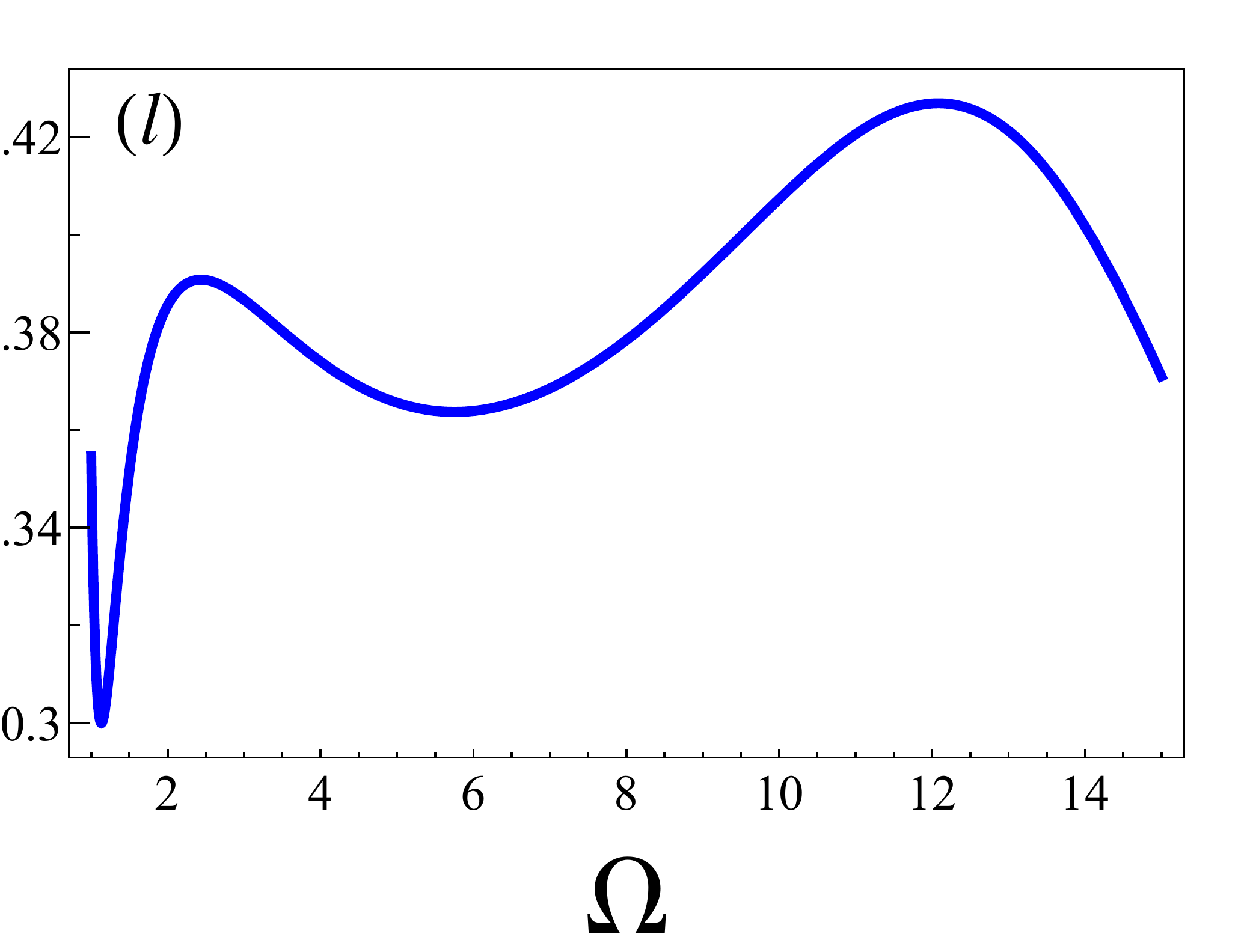}    
        \end{subfigure} 
\end{center}
\caption{ Response $A/A_{0}$ as a function of angular frequency ($\Omega $)
of the periodic external driving force as given in Eq.~(\ref{resp07}).
Maximum of $A/A_{0}$ for specific parameters of the system describes a
resonance between the behavior of $x$ and the external drive. \textbf{(g)} $%
\gamma /m=0.2$, $\omega ^{2}/m=1$, $\lambda _{1}=0.1$%
, $\lambda _{2}=0.1$, $\sigma _{1}=0.1$, $\sigma %
_{2}=0.1$. \textbf{(h)} The same parameters as in \textbf{(g)}, except that $\sigma_{1}=0.7$. 
\textbf{(i)} The same parameters as in \textbf{(h)}, except that $\sigma_{2}=0.95$. 
\textbf{(j)} The same parameters as in \textbf{(i)}, except that $\lambda_{1}=10$. 
\textbf{(k)} The same parameters as in \textbf{(i)}, except that $\lambda_{2}=10$. 
\textbf{(l)}The same parameters as in \textbf{(i)}, except that $\sigma_{1}=0.995$ and $\sigma_{2}=0.75$. 
}
\label{figresp03}
\end{figure}

\subsubsection{Random Mass and Damping}

When both sources of noise (random mass and random damping) are present, we
expect that a mixture of the previously discussed cases to take place. In
panel \textbf{(g)} of Fig.~\ref{figresp03},  $A/A_{0}$ exhibits a resonance
for specific $\Omega $, while the strengths of the sources of noise are
quite small, $\sigma _{1}=0.1$ and $\sigma _{2}=0.1$. Increasing the
strength of the random mass noise, while leaving the strength of the random
damping noise constant, splits the resonance. Panel \textbf{(h)} of Fig.~\ref%
{figresp03} shows two maxima for $A/A_{0}$ and the effect is similar to the
case of only a random mass, as described in panel \textbf{(b)} of Fig.~\ref%
{figresp01}. The presence of a small noise term for the damping does not
qualitatively change the effect. But if in addition to increasing the
strength of $\xi _{1}$, one also increase the strength of $\xi _{2}$ ( i.e.
random damping), a non-symmetric effect occurs. For the case of only random damping,
an increase of noise strength expands the size of the resonance (panel 
\textbf{(e)} of Fig.~\ref{figresp02}). In panel \textbf{(i)} of Fig.~\ref%
{figresp03}, we see that as the strength of $\xi _{2}$ increases, it does not
affect the values of the maxima in the same fashion. While the second maxima
that appeared in panel \textbf{(h)} of Fig.~\ref{figresp03} expanded
significantly , the first maxima grew only  slightly. 
This asymmetry arises due to the asymmetry of the resonant frequencies (as function of mass and damping) at each intrinsic state of the oscillator (i.e. Eq.~(\ref{simplestate})). $m$ appears in the denominator  and affects $\Omega_R$ more violently than $\gamma$ that appears in the numerator. Due to this fact a significant effect is expected for the state with smaller $m$ and small $\gamma$. 
The temporal correlation must be long enough in order to observe the mentioned effect and indeed increasing either $\lambda_1$ 
(panel \textbf{(j)} of Fig.~\ref{figresp03}) or $\lambda_2$ (panel \textbf{(k)} of Fig.~\ref{figresp03}) reverses the response to previously observed cases.

\begin{figure}[t]
\begin{center}
\begin{subfigure}[b]{0.3755\textwidth}
                \includegraphics[width=\textwidth]{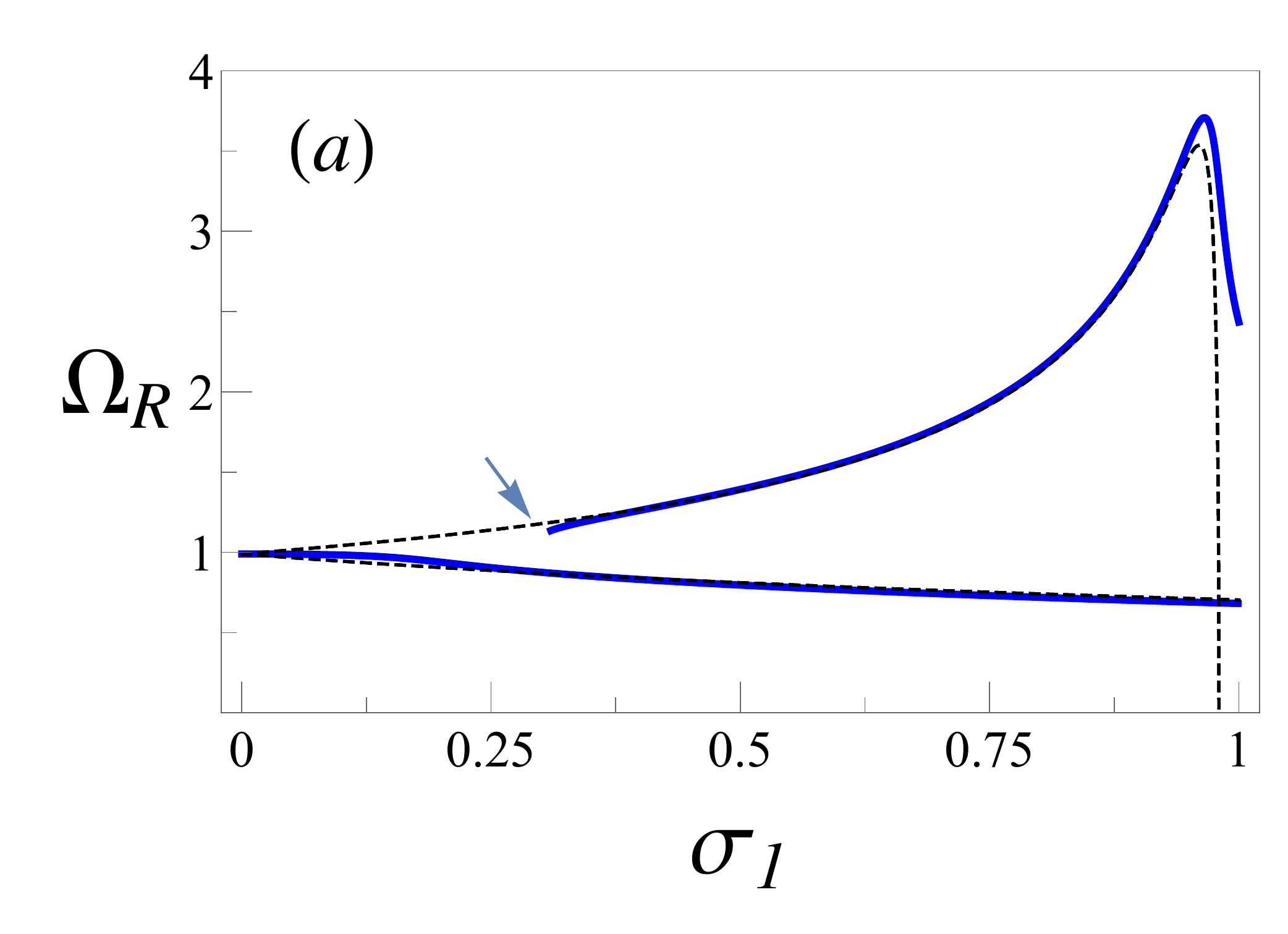}
              
        \end{subfigure} ~ 
        \\
\begin{subfigure}[b]{0.3755\textwidth}
                \includegraphics[width=\textwidth]{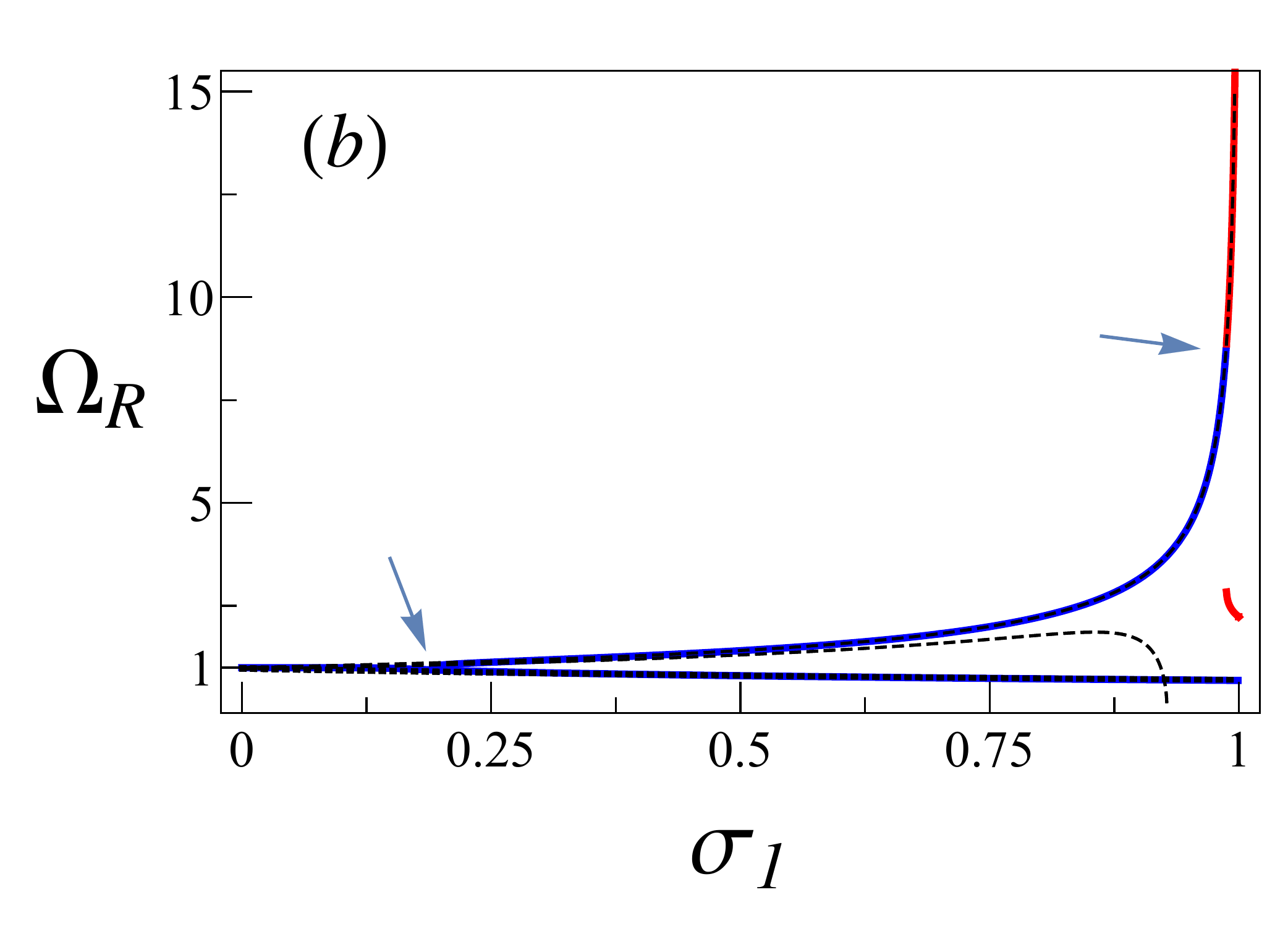}
              
        \end{subfigure} ~ 
\begin{subfigure}[b]{0.3755\textwidth}
                \includegraphics[width=\textwidth]{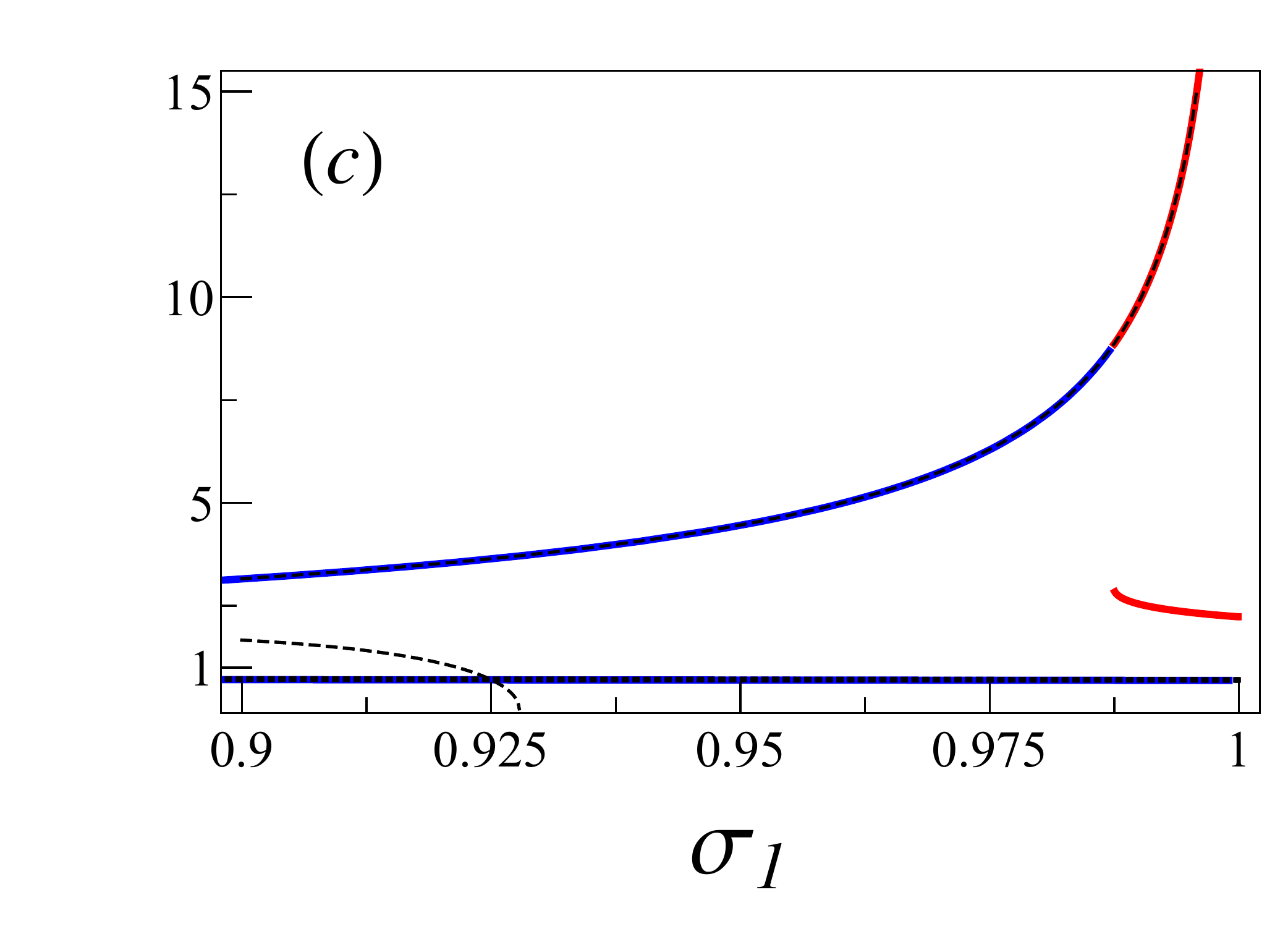}    
        \end{subfigure}
\end{center}
\caption{ Resonant frequencies $\Omega_R$ as function of $\xi_1$ strength, for the case of random mass and random damping (thick lines) and for the specific states of the oscillator (dashed lines).  \textbf{(a)} Only random mass noise is present $\gamma/m=0.2$, $\omega^2/m=1$, $\lambda_1=0.1$, $\lambda_1=0.0$,$\sigma_2=0$. The arrow points to an emergence of a second resonance.  \textbf{(b)} Both noises are present.  $\gamma/m=0.2$, $\omega^2/m=1$, $\lambda_1=0.1$, $\lambda_1=0.1$,$\sigma_2=0.9$. The left arrow shows the position of emergence of the second resonance while right arrow to the position of emergence of the third resonance. Four different dashed lines are presented, two bottom lines almost coincide for the whole range of $\sigma_1$. The different color of the thick line is plotted for the part when three resonances occur. \textbf{(c)} Zoom into the range $0.9\leq\sigma_1\leq1$ of panel \textbf{(b)}. }
\label{fig06}
\end{figure}

In the case of random damping, the presence of two states does not lead to appearance of resonant splitting. Interestingly enough, when both random damping and random mass present, an additional resonance splitting can occur. 
By keeping the temporal correlation of both sources of noise sufficiently long $\lambda
_{1}=\lambda _{2}=0.1,$ we take the limit of very large strength of a random
mass noise ($\sigma _{1}=0.995$) and large strength of random damping noise (%
$\sigma _{2}=0.7$). The result of additional resonance is presented in panel \textbf{(l)} of Fig.~\ref{figresp03}. Obviously,  
the simplistic approach that describes each resonant frequency as a frequency that correspond to a resonance for one of the states of the oscillator, fails here. 

In order to study this effect further we present the behavior of the resonant frequency $\Omega_R$. 
In Fig.~\ref{fig06} panel {\bf(a)} the behavior of the resonant frequency is presented for the case of random mass without random damping and compared to the predictions of Eq.~(\ref{simplestate}). The second resonant frequency appears only when the frequencies of the two states are sufficiently distinct, and in general the behavior of the noisy case follows the predictions for the two different states. Even the non-monotonicity of $\Omega_R$ for random mass is a consequence of the non-monotonicity of $\Omega_R$ in Eq.~(\ref{simplestate}). When also random damping is present the situation is quite similar while $\sigma_1$ is small enough. In panel {\bf(b)} of Fig.~\ref{fig06}  behavior similar to panel {\bf(a)} is presented. The four different states appear as two states (the dashed lines are very close to one another) and generally there is almost no obvious effect of the additional noise. For large enough $\sigma_1$ Eq.~(\ref{simplestate}) predicts disappearance of resonance for one of the states of the oscillator (one of the dashed lines drops to zero). Inside this region where only two states with resonance exist suddenly appears additional resonance for the noisy case (lower red line). We cannot attribute this resonance to a resonance in an intrinsic state of the oscillator, since this intrinsic resonance does not exist for this range of parameters.

While for majority of the cases we managed to describe the response behavior in terms of response of the intrinsic states of 
the oscillator, there are exceptional situations. In those situations appearance of an additional resonance must be interpreted as an interference between various intrinsic states of the oscillator and not an attribute of a response in a single state. 
The noises in our oscillator model are not only capable of creating an intrinsic state that will attain a proper response. An effective coupling between transitions manages to create a preferable response to an external filed. Further study of such coupling is needed.


\section{Conclusions}

We considered an oscillator with two multiplicative random forces, which
define the random damping and the random mass. The random mass means that
the molecules of the surrounding medium, not only collide with an
oscillator, but also adhere to it for a random time, thereby changing  the
oscillator mass. We calculated the first and the second moments of the
oscillator coordinate by considering these two moments in the form of the
damped exponential functions of time, $\exp (\alpha t).$ The signs of $\alpha $ , which are obtained numerically, define the mean and energetic
stability of the system. Stable solutions of the moments were represented by
determinants of appropriate matrices. We brought references to many
applications of such calculations to physics, chemistry and biology. 
Specifically we have shown that for the mean stable oscillations persist at the transition to instability.

The
last section described the stochastic resonance phenomenon, that is  the
noise increased the applied periodic signal
by helping the system to absorb more energy from the external force~\cite%
{Yung} . We presented the stochastic resonance as the function of the
frequency $\Omega $ of the applied periodic signal, first separately for a
random mass and random damping, and for the case of joint action of both
these sources of  noise. For most cases we managed to  describe the observed phenomena in terms of simple intrinsic states 
of the oscillator and presence/non-presence of resonance for those states. Description by the means of underlying 
intrinsic states might become useful in experimental situations where the intrinsic states are explored by the means of response to external field, e.g., biomolecule folding/unfolding experiments~\cite{unfolding01,unfolding02} where distinct folded/unfolded sates are explored by external pulling .
While the description by the means of response of the intrinsic states holds for majority of the cases, we found exceptions to this simple description. Specifically, we argue that appearance of additional resonant frequency at a regime where intrinsic resonance frequency dies out occurs due to transitions between states and not a presence 
of a single preferable response in an intrinsic state. It is the regime where the interference between states creates a preferable response.   

\begin{figure}[t]
\begin{center}
\begin{subfigure}[b]{0.3655\textwidth}
                \includegraphics[width=\textwidth]{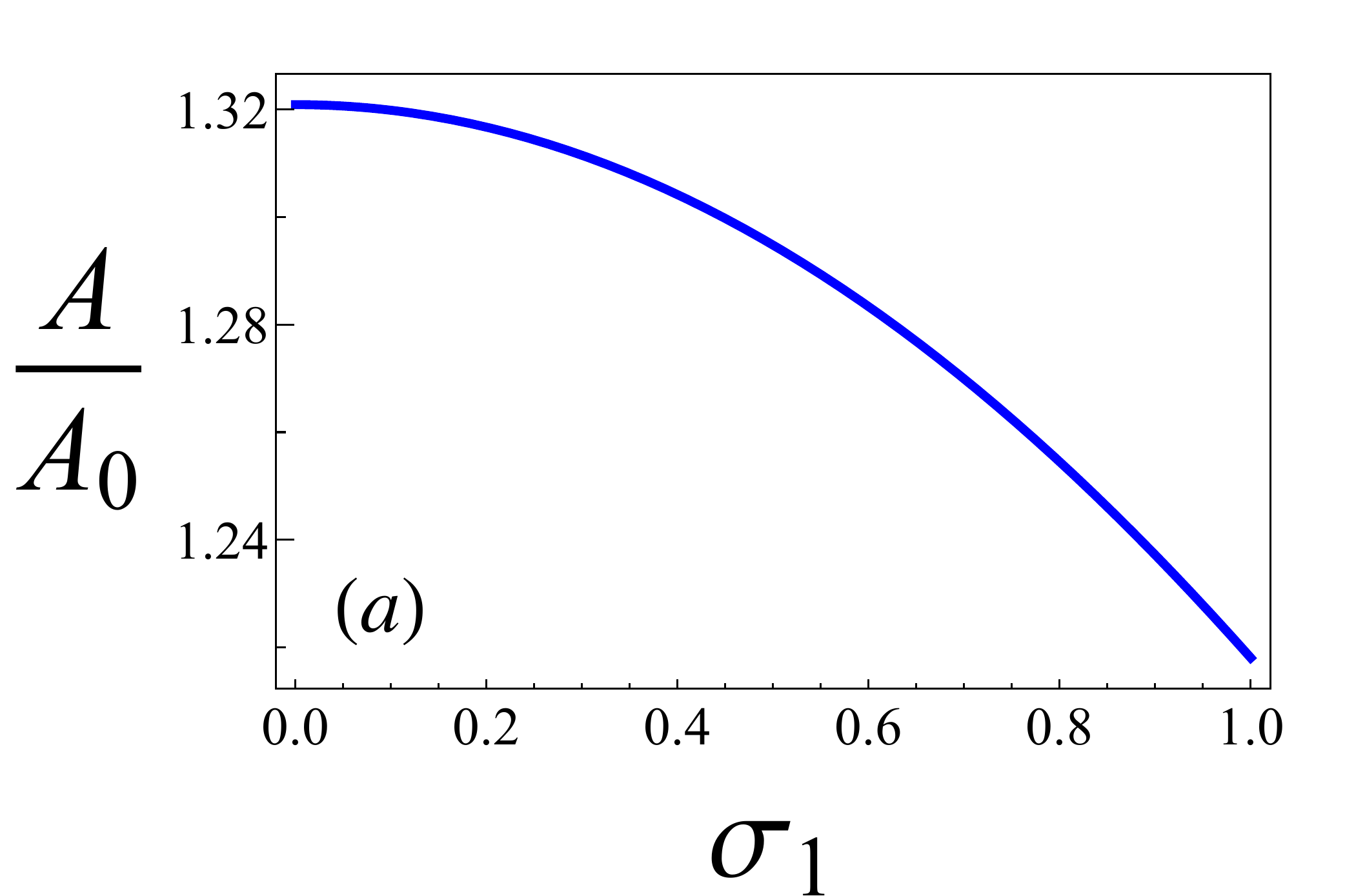}
              
        \end{subfigure} ~
        \begin{subfigure}[b]{0.3755\textwidth}
                \includegraphics[width=\textwidth]{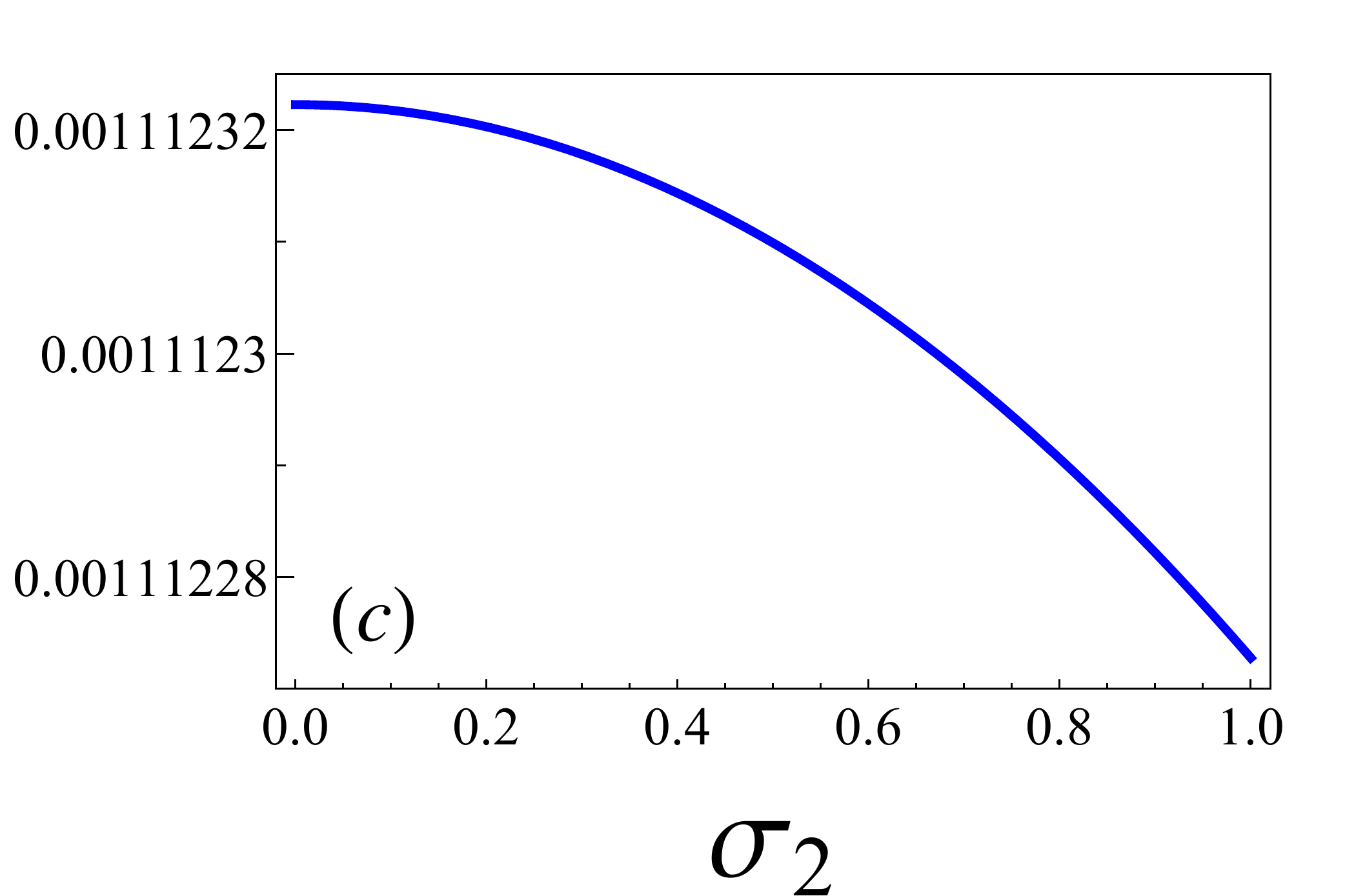}
              
        \end{subfigure}
        \\
\begin{subfigure}[b]{0.3655\textwidth}
                \includegraphics[width=\textwidth]{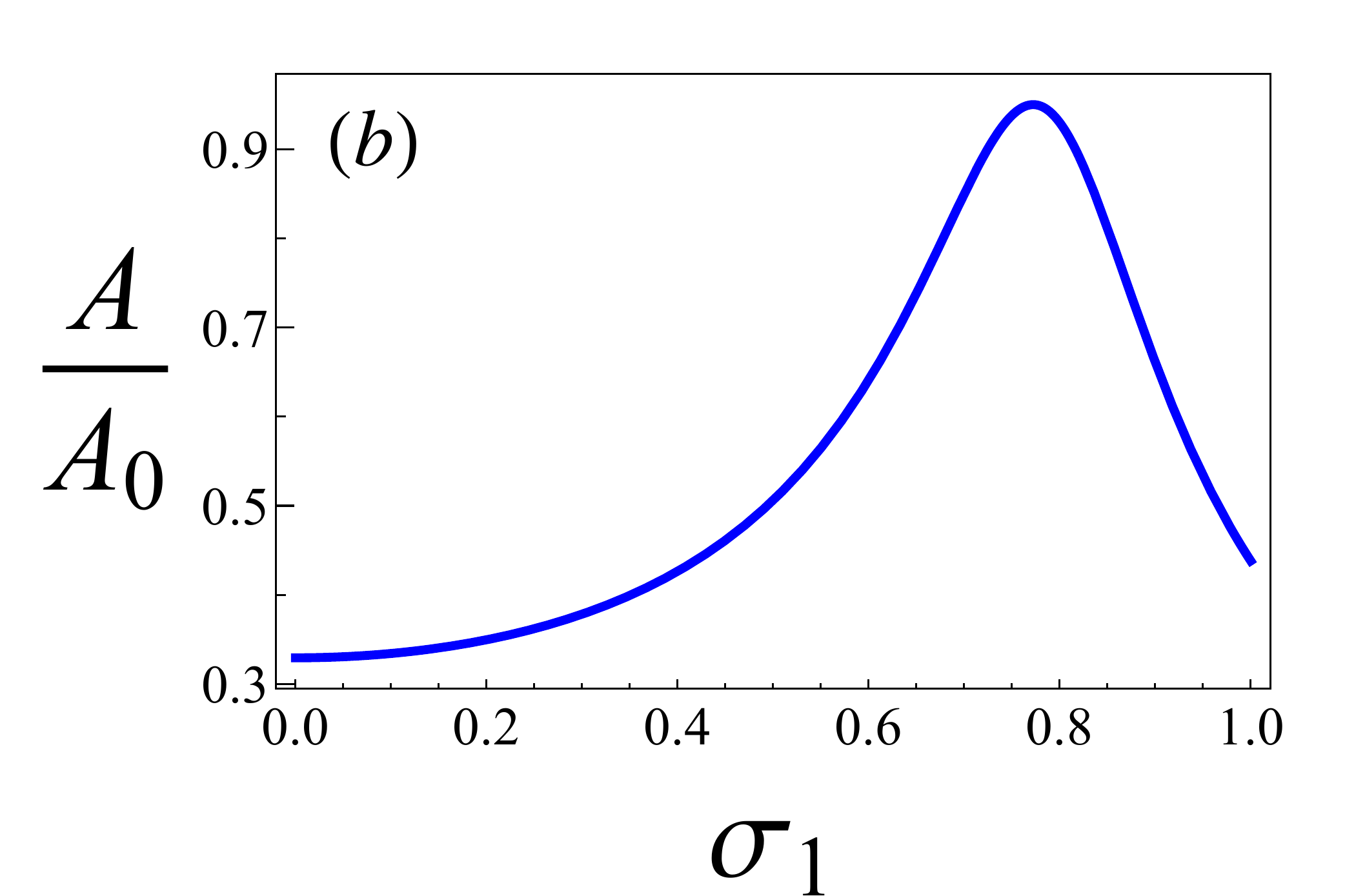}
              
        \end{subfigure} ~ 
\begin{subfigure}[b]{0.3755\textwidth}
                \includegraphics[width=\textwidth]{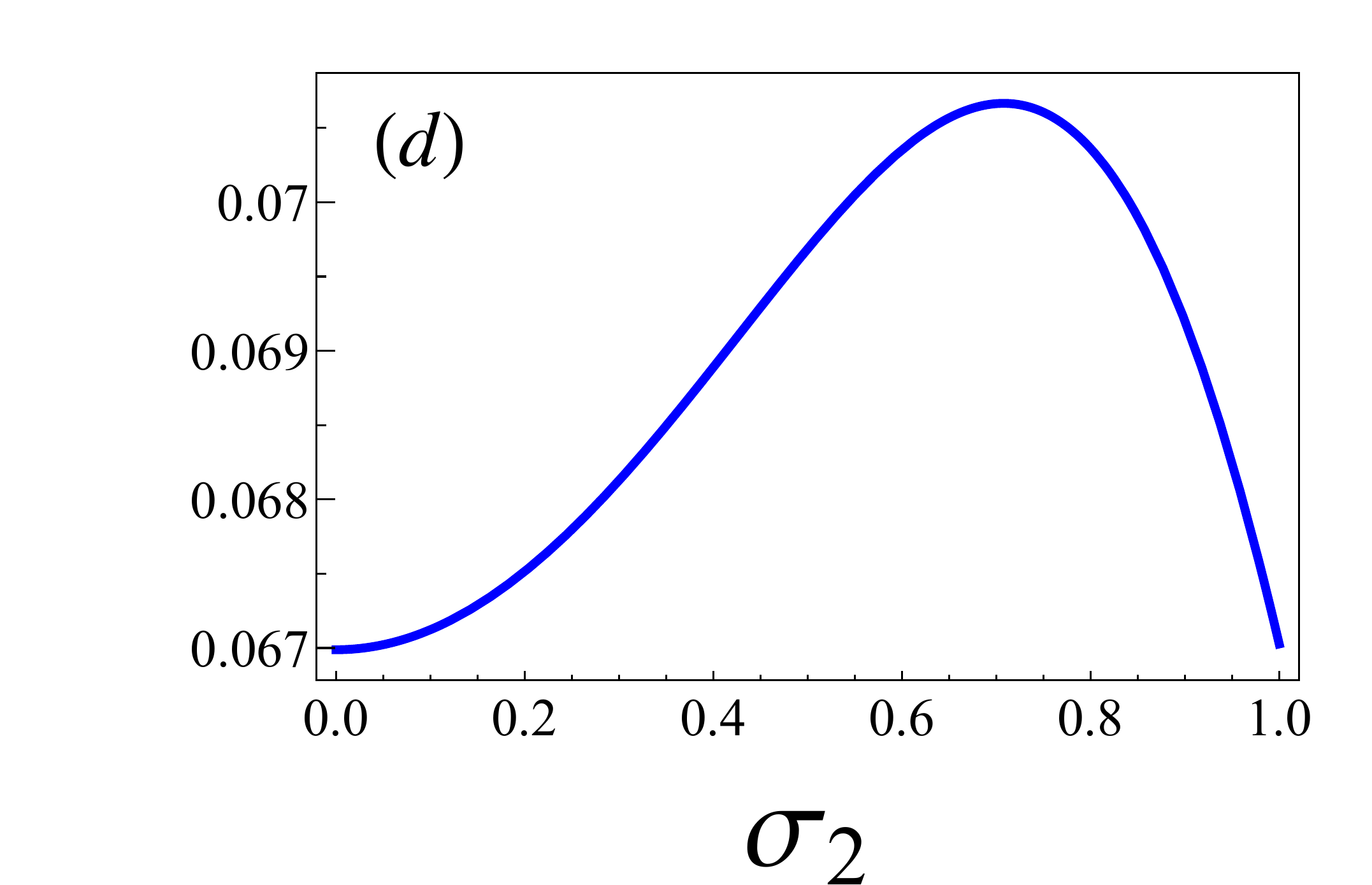}    
        \end{subfigure}
\end{center}
\caption{ Response $A/A_{0}$ as a function of noise strenght ($\sigma_1$ and $\sigma_2$)
of the periodic external driving force as given in Eq.~(\ref{resp07}) for spcific values of $\Omega$.
 \textbf{(a)} $\gamma /m=0.2$, $\omega ^{2}/m=1$, $\lambda _{1}=0.5$%
, $\lambda _{2}=0.5$, $\sigma _{2}=0.5$, $\Omega=0.5$. 
\textbf{(b)} The same parameters as in \textbf{(a)}, except that $\Omega=2$. 
\textbf{(c)} $\gamma /m=0.2$, $\omega ^{2}/m=1$, $\lambda _{1}=0.5$%
, $\lambda _{2}=0.5$, $\sigma _{1}=0$, $\Omega=30$. 
\textbf{(d)} The same parameters as in \textbf{(c)}, except that $\sigma_1=0.994$. 
 }
\label{figapp01}
\end{figure}

\section{Appendix}

In the main text we presented the response $A/A_0$ as a function of $\Omega$. In this Appendix we present the response as a function of noise strength $\sigma_1$ and $\sigma_2$. In general the dependence of $A/A_0$ on the noise strength, for  specific value of $\Omega$, is associated with the chosen $\Omega$. Non-monotonic behavior is expected in regions of $\Omega$ where the resonant frequency $\Omega_R$ will be shifted when changing the noise strength ($\sigma_1$ or $\sigma_2$). If $\Omega_R$ will coincide with the chosen $\Omega$ for some value $0<\sigma_1<1$ (or $\sigma_2$) a maxima of $A/A_0$ will appear for this specific value of $\sigma_1$ (or $\sigma_2$). When such crossover doesn't occur the behavior of $A/A_0$ is monotonic as displayed in Fig.~\ref{figapp01} panels (a) and (c). When a crossover of $\Omega_R$ occurs a modest maxima will be observed, as described in panels (b) and (d). Appearance of maxima as a function of $\sigma_2$ occurs for non zero values of $\sigma_1$. In the main text we described situations, when both $\sigma_1$ and $\sigma_2$ are non-zero, where two maxima of $A/A_0$ appear (as a function of $\Omega$). Existence of two (or even three) $\Omega_R$ suggest that when those resonant frequencies are shifted one might observe also two maxima for $A/A_0$ as function of the noise strength. Due to the fact that the maxima of $A/A_0$ (as function of $\Omega$) are well separated (in $\Omega$) we were unable to find parameters where this phenomena might occur.

\bibliographystyle{ieeetr}
\bibliography{./oscill01.bib}

\end{document}